\documentclass{aastex}

\usepackage[onecolumn]{emulateapj5}
\usepackage{apjfonts}

\shortauthors{RICKER \& SARAZIN}
\shorttitle{OFF-AXIS CLUSTER MERGERS}
\submitted{Accepted by ApJ June 22, 2001}

\begin{document}

\title{Off-Axis Cluster Mergers: Effects of a Strongly Peaked
	Dark Matter Profile}

\author{Paul M. Ricker}

\affil{Department of Astronomy and Astrophysics, University of Chicago,
	5640 S. Ellis Ave., Chicago, IL 60637; ricker@flash.uchicago.edu}

\and

\author{Craig L. Sarazin}

\affil{Department of Astronomy, University of Virginia, 
	P.O. Box 3818, Charlottesville, VA 22903-0818;
	sarazin@virginia.edu}

\begin{abstract}

We present a parameter study of offset mergers
between clusters of galaxies. Using the Eulerian
hydrodynamics/$N$-body code COSMOS, we simulate mergers between
nonisothermal, hydrostatic clusters with a steep central dark matter density
profile and a $\beta$-model gas profile.
We constrain global properties of the model
clusters using observed cluster statistical relationships.
We consider
impact parameters between zero and five times the dark matter scale radius
and mass ratios of 1:1 and 1:3.
The morphological changes, relative velocities, and temperature jumps
we observe agree with previous studies using the King profile
for the dark matter.
We observe a larger jump in X-ray luminosity ($\sim 4-10\times$)
than in previous work,
and we argue that this increase is most likely a lower limit due to our
spatial resolution.
We emphasize that luminosity and temperature
jumps due to mergers may have an important bearing on constraints on $\Omega$
derived from the observation of hot clusters at high redshift.
Shocks are relatively weak in the cluster cores;
hence they do not significantly increase the entropy there.
Instead, shocks create entropy
in the outer regions, and this high-entropy gas is mixed with
the core gas during later stages of the merger.
Ram pressure initiates mixing by
displacing the core gas from its potential center, causing it to become
convectively unstable. The resulting convective plumes produce large-scale
turbulent motions with eddy sizes up to several 100~kpc. This turbulence
is pumped by dark matter-driven oscillations in the gravitational potential.
Even after nearly a Hubble time these motions persist as subsonic
turbulence in the cluster cores, providing $5-10\%$ of the support
against gravity.
The dark matter oscillations are also reflected in the extremely long
time following a merger required for the remnant to reach virial equilibrium.

\end{abstract}



\section{Introduction}
\label{Sec:introduction}

The intracluster medium (ICM) in clusters of galaxies is now understood
to evolve significantly over the lifetime of a cluster due to the complex
interactions induced by mergers with other clusters.
These mergers are an important part of the cluster formation process in
present hierarchical models of large-scale structure.
Mergers supply a substantial amount of energy to a cluster
(two $10^{15}M_\odot$ clusters colliding at $\sim 1000$~km~s$^{-1}$ yield
an energy $\sim 10^{64}$~erg).
Thus they have been suspected of contributing to many different cluster
phenomena, including the production of radio halos
(Harris et al.\ 1980; Burns et al.\ 2000),
extreme ultraviolet and nonthermal X-ray emission
(Sarazin \& Lieu 1998; Blasi \& Colafrancesco 1999; Blasi 2000),
the bending of narrow-angle-tail and wide-angle-tail radio
galaxies (Burns et al.\ 1994; G\'omez et al.\ 1997; Bliton et al.\ 1998),
and the disruption of cooling flows
(McGlynn \& Fabian 1984; Stewart et al.\ 1984).

While some evidence for mergers can be obtained from optical data alone,
it is the X-ray-emitting gas which provides the strongest evidence
for mergers.
The energy gained by the ICM of one cluster as it falls into the potential
well of another is easily enough to raise its temperature to the observed
values of $kT\sim 10^7 - 10^8$~K.
Simulations have shown that
the resulting shocks can produce long-lived substructure in X-ray surface
brightness maps; however, they have also shown that X-ray temperature and
velocity maps are likely to be much better diagnostics, partly
because shock brightness contrasts are small when seen in projection
(Roettiger, Burns, \& Loken 1996; Ricker 1998).
The {\it Chandra} and {\it XMM} satellites are producing ICM temperature maps
of much higher spatial and spectral resolution and energy sensitivity than
was previously possible
(e.g., Markevitch et al.\ 2000;
Vikhlinin, Markevitch, \& Murray 2001;
Mazzotta et al.\ 2001).
While generally confirming that mergers produce large temperature variations
in clusters, these new observations have raised new questions.
In addition to showing high-temperature ridges characteristic of shocks,
many of these observations show cold cluster cores moving through low
density, shock-heated intracluster gas.

The intracluster gas is repeatedly stirred during a cluster's life by
merger shocks (Norman \& Bryan 1998) and galaxy wakes
(Stevens, Acreman, \& Ponman 1999; Sakelliou 2000).
The Reynolds number $Re$ associated with these motions is typically of the
order (Sarazin 1988)
\begin{equation}
Re \approx 3{\cal M}\left({\ell\over\lambda_i}\right)\ ,
\end{equation}
where ${\cal M}$ is the Mach number, $\ell$ is the size of the stirring agent,
and $\lambda_i \sim {\rm kpc}$ is the mean free path for ions in the ICM.
Merger shocks (${\cal M}\sim 1-4$) typically span lengths comparable to
the size of a cluster ($\ell\sim {\rm Mpc}$), so for them $Re\sim 4\times10^3$.
Galaxy motions ($\ell\sim 30{\rm\ kpc}$)
are transonic relative to the ICM, yielding $Re\sim 10^2$.
These influences on the ICM may produce turbulence; if so, it is in a
range of Reynolds number which is accessible to present simulations.
However, definitive observational detection of turbulence in the ICM 
will require measurements of the local gas velocity field.
X-ray observations are just beginning to enter the regime in which
it will be possible to constrain these velocities using
X-ray emission lines (Roettiger \& Flores 2000).

Turbulence may contribute to a number of the proposed
secondary consequences of mergers, including the formation of
constant-density gas cores, the destruction of cooling flows, and
{\it in situ} acceleration of cosmic-ray particles.
It can also affect
the stripping of gas from galaxies as they orbit in the cluster potential.
Unless tangled small-scale magnetic fields limit diffusion
(Chandran et al.\ 1999),
this picture is further complicated by the breakdown of the fluid description
of the ICM on kpc scales.
Nevertheless, we can make some headway with this problem by asking
whether merger-induced hydrodynamical effects on scales of tens to
hundreds of kpc can feed energy to scales on which collisionless processes
become important. It may also be possible to include collisionless transport
effects as a correction to fluid calculations on small scales.

Offset cluster mergers are a likely source of vorticity and turbulence
in the intracluster medium.
The bulk angular momentum that feeds this vorticity originates in tidal
torques produced during the linear phase of structure formation
(Hoyle 1949; Peebles 1969; White 1984)
and should result in mergers with nonzero impact parameters.
One example of such an off-center merger is Abell~754
(Zabludoff \& Zaritsky 1995; Henriksen \& Markevitch 1996;
Roettiger, Stone, \& Mushotzky 1998; Valinia et al.\ 1999).

To distinguish the effects of offset mergers from those of the many other
simultaneous influences on the ICM, it is necessary to study them in isolation
(see Schindler [2000] for a brief review).
Offset mergers have been studied by only a few groups.
Roettiger et al.\ (1998) and Roettiger \& Flores (2000) treat 
them primarily in the context of attempts to model specific observed
clusters.
Ricker (1998) performed a parameter study of offset mergers between
equal-mass clusters but neglected the effects of the collisionless dark
matter component thought to dominate the mass of most clusters.
Takizawa (2000) considers electron-ion nonequipartition effects in a
parameter study of offset mergers, but these smoothed-particle hydrodynamics
calculations use too few particles to adequately characterize merger shocks
and energy transfer between the gas and dark matter.
Also, published studies of controlled mergers have used dark matter density
profiles with constant-density cores, despite the fact that simulations of
hierarchical structure formation produce halos with cuspy central density
profiles (e.g., Navarro, Frenk, \& White 1997 [NFW]; Moore et al.\ 1998;
Jing \& Suto 2000).
Finally, none of these systematic studies of controlled mergers emphasizes
the distribution of entropy or the development of turbulence.

In this paper we describe a parameter study of offset mergers
that addresses these issues. We ask the following questions.
Can constant-density gas cores survive in collisions between clusters
with steep central total density profiles?  How long can the pressure
peaks associated with these cores survive in off-center collisions?
How do typical offsets affect the luminosity and temperature jumps
expected during mergers?
How is entropy generated and redistributed in mergers?
How much do merger-driven motions mix the cluster cores and contribute to
nonthermal support of the gas?

In \S~\ref{Sec:numerical} we describe the simulation code and isolated
merger model we use in our calculations.
Our single-cluster convergence study and the merger simulations
themselves are described in \S~\ref{Sec:simulations}.
In \S~\ref{Sec:discussion}, we discuss the structure of the merger remnants,
the evolution of luminosity and temperature,
and the survival of distinct pressure peaks in the simulations.
Finally, \S~\ref{Sec:conclusions} presents our conclusions.
We will discuss at greater length merger-induced turbulence and
methods for detecting off-center collisions in a future paper.

When cosmological scaling is required, in this paper we have assumed a
Hubble constant $H_0 = 100h$~km~s$^{-1}$~Mpc$^{-1}$ with $h=0.6$.
Unless otherwise specified, quantities are expressed using length units
of $h^{-1}{\rm\ Mpc}$, mass units of $10^{15}M_\odot$, time units of Gyr,
and temperature units of keV.


\section{Numerical Method}
\label{Sec:numerical}

\subsection{Simulation Code}

For our merger simulations we have used the COSMOS $N$-body/hydrodynamics 
code (Ricker, Dodelson, \& Lamb 2000).
COSMOS solves the Euler equations for the intracluster medium using the
piecewise-parabolic method (PPM; Colella \& Woodward 1984).
The code follows the evolution of collisionless dark
matter by means of the particle-mesh method
(Hockney \& Eastwood 1988), solving the Poisson equation for the total
gravitational potential using a full multigrid algorithm (Brant 1977).
It also incorporates radiative cooling of the gas, although the simulations
described in this paper do not make use of this feature.
All three of the primary code components (hydrodynamics, gravity, and
dark matter) make use of static, nonuniform grids, allowing us to use
smaller zones in regions of interest than would be possible with a
uniform-grid code.
Our code is fully compressible and uses a three-dimensional Cartesian grid,
so shocks and convective effects are intrinsically included.
COSMOS has been tested thoroughly against a suite of standard test
problems, and we have already used its hydrodynamical
and gravitational modules to study cluster mergers without dark matter
(Ricker 1998).

The combination of PPM and particle-mesh is widely used in cluster
dynamics and large-scale structure simulation.
PPM is to be contrasted with
another widely-used hydrodynamical method, smoothed particle hydrodynamics
(SPH; Gingold \& Monaghan 1977; Lucy 1977), a Lagrangian method
which uses particles as
moving interpolation centers for the hydrodynamical variables.
We use PPM because it resolves shocks extremely well. SPH handles shocks
poorly but compensates in part by achieving better overall spatial
resolution in high-density regions.
Both algorithms have been shown to give
consistent results when run with the same cluster-formation initial
conditions (Frenk et al.\ 1999).


\subsection{Boundary Conditions}

Clusters formed in cosmological large-scale structure
calculations do not evolve in isolation.
Generally they are subject to tidal torques in the linear phase of
structure growth, and during the nonlinear phase they accrete
a significant number of smaller objects, from galaxies to other clusters.
However, we are interested here in separating out the effects of
merger-generated shocks.
Because we are studying controlled mergers in isolation, it makes sense
to use outflow boundary conditions for the gas and dark matter
and isolated boundaries for the gravitational field.
As noted in \S~\ref{Sec:cluster scaling}, this requires us to
cut off the initial density profiles at a finite radius, allowing particles
and gas to escape through this outer boundary as the simulation progresses.
Our single-cluster convergence study (\S~\ref{Sec:convergence})
shows that this leads to a steepening of density profiles at large radius
which is unrelated to the merger-induced changes in the clusters.
For this reason we confine our conclusions to the region within
the initial cutoff radius, focusing particularly on the behavior of the
clusters within the innermost few core radii of their centers.

Here we briefly describe our implementation of isolated boundary conditions;
further details appear in Ricker et al.\ (2000).
For hydrodynamics
we use standard zero-gradient boundary conditions wherever the mass flux
through the external boundary is positive. For edge zones with inward-directed
velocities, we use zero-gradient boundary conditions for all variables except
the component of velocity normal to the boundary; for this component we use
a Dirichlet boundary condition. This procedure prevents the destabilizing
artificial inward flows which can result when matter near the boundary begins
to fall back toward the center.
In the case of dark matter particles we
discard those particles which leave the grid.
To handle isolated boundaries with our Poisson solver we have developed a
method which first obtains a zero-boundary solution, uses this to obtain
boundary values for the required correction to this solution, then solves for
the correction using the given boundary values.


\subsection{Initial Cluster Model}
\label{Sec:initial}

\subsubsection{Density Profiles}
\label{Sec:initial_profiles}

In this paper we approach the merger problem by simulating isolated collisions
between idealized model clusters.
Our model clusters are initially spherically symmetric, with specified
density profiles for the gas and dark matter.
For the questions we address, this approach has two main advantages over
studying clusters formed in large-scale structure simulations.
We eliminate the complication of multiple mergers and focus on the
effects of a single collision, allowing us to isolate important physical
mechanisms.
We also achieve high spatial resolution, which is critical for obtaining
suitably converged results that track the length scales of interest
in cluster cores.
However, this approach does not consider the effects of
multiple simultaneous mergers, external tidal fields, and infalling matter.
Additionally, we neglect radiative cooling and choose our cluster parameters
to ensure that the cooling timescale is larger than a Hubble time.
We do this to avoid the computational difficulties that arise when the gas
is allowed to cool very quickly (the so-called `cooling catastrophe').
The effects of mergers on cooling flows present some very interesting and
challenging problems, which we will take up in a future paper.

In the current picture, the formation of a cluster of galaxies is a process of
steady gravitational
accretion and relaxation toward equilibrium, periodically interrupted by
mergers with other partially virialized systems. These periodic interruptions
destroy any pre-existing equilibrium. After such a merger, the cluster
resumes relaxing toward a new equilibrium state, continuing until it is again
interrupted. Previous calculations have shown that this relaxation
time $t_{\rm relax}$
is long compared to the likely merger interval $t_{\rm merge}$
(Roettiger et al.\ 1993; Schindler \& M\"uller 1993; Pearce, Thomas,
\& Couchman 1994; Ricker 1998),
meaning that
mergers usually involve clusters which are not fully virialized. However,
studying the case of large $t_{\rm merge}/t_{\rm relax}$ is a necessary
preliminary to study of the case in which a merger perturbs a partially
virialized initial state.
In this paper we are interested in understanding the physical mechanisms
important in individual mergers rather than simulating a specific observed
cluster.

Accordingly, we begin with an initial model that represents our best current
knowledge of the structure of fully virialized, noninteracting clusters.
We assume functional forms for the gas density and total density profiles,
then set the dark matter profile, the gas temperature profile, and the
dark matter velocity dispersion profile in such a way as to establish
hydrostatic equilibrium.
The gas density profile is taken from the $\beta$-model (Cavaliere
\& Fusco-Femiano 1976),
\begin{equation}
\label{Eqn:beta model}
\rho_g(r) = \rho_{g0}\left[1+\left({r\over r_c}\right)^2\right]^{-3\beta/2}\ ,
\end{equation}
where the central density $\rho_{g0}$, core radius $r_c$, and asymptotic
slope parameter $\beta$ are fitting parameters.
When used with the assumption of isothermality, the $\beta$-model
with $\beta\approx 2/3$ produces
an X-ray surface brightness profile which is consistent with those observed
in many clusters without cooling flows (Jones \& Forman 1984).
In this paper we take $\beta=2/3$ as a representative value.
The $\beta$-model is
similar in shape to the nonsingular isothermal sphere profile (Binney \&
Tremaine 1987), which is a solution of the gas hydrostatic equation under the
assumption of isothermality.

We take guidance for the form of the dark matter density profile
from models of hierarchical large-scale structure formation.
Using $N$-body simulations, NFW have shown
that, in such models,
dark matter halos on scales from individual galaxies to rich clusters form
with density profiles given by
\begin{equation}
\label{Eqn:nfw model}
\rho(r) = \rho_s\left[\left({r\over r_s}\right)\left(1+{r\over r_s}\right)^2
        \right]^{-1}\ ,
\end{equation}
where $\rho_s$ and $r_s$ are, respectively, a scaling density and radius,
both of which depend upon the mass of the halo in question. The mass dependency
in these simulations arises because of the differing redshift of formation
of halos with differing masses, with the sense of the dependency being that
less massive halos, which form earlier, have higher central densities and
are thus more concentrated. The exact shape of this universal density profile
has been called into question by Moore et al.\ (1998), who favor a steeper
profile near the centers of halos, arguing that previous simulations have
insufficiently resolved the halos. Nevertheless, the clear prediction of
these models is that self-similar, purely gravitational evolution leads to
a steeper density profile than that predicted by the $\beta$-model.
The addition of collisional, self-gravitating gas most likely modifies this
profile somewhat, though if dark matter dominates as we expect it to,
the total density profile should still be steeper than that of the gas near
the center.
Therefore, we choose a total density profile using the NFW functional
form. The dark matter
density profile is given by the difference between the total profile and
the gas profile:
\begin{equation}
\rho_{dm}(r) = \rho(r) - \rho_g(r)\ .
\end{equation}
This can only be an approximation to the true situation, because
the NFW profile has been established by $N$-body simulations which
do not include gasdynamics.
If the NFW profile were to be maintained for the total density upon
including gasdynamics, the shape of the dark matter profile would need
to be different enough to exactly compensate for the $\beta$-model shape
of the gas. However, because the dark matter dominates the total density,
particularly in the center of the cluster where the NFW profile continues
to rise while the $\beta$-model tends to a constant value, this is most
likely an acceptable approximation in the inner regions of non-cooling-flow
clusters.

\subsubsection{Temperature and Velocity Dispersion Profiles}
\label{Sec:temperature profiles}

We choose the gas temperature and
dark matter velocity dispersion profiles from the equations of hydrostatic
equilibrium for the two matter components.
A more complete discussion will be given in a subsequent paper
(Ricker \& Sarazin 2001, in preparation).
With the total density profile given by equation (\ref{Eqn:nfw model}),
the total mass $M$ enclosed within radius $r$ is given by
\begin{equation}
\label{Eqn:enclosed mass}
M(r) = 4\pi\rho_s r_s^3\left[\ln(1+x) - {x\over 1+x}\right]\ ,
\end{equation}
where $x\equiv r/r_s$.
The hydrostatic equation for the gas temperature profile $T(r)$ is
\begin{equation}
{d\over dr}\left[{\rho_g(r)kT(r)\over \mu m_H}\right] =
	-{GM(r)\over r^2}\rho_g(r)\ ,
\end{equation}
where $\mu m_H$ is the average mass per particle in the gas.
It is useful to define a characteristic temperature scale $T_s$ as
\begin{equation}
\label{Eqn:temperature scaling}
T_s \equiv {4\pi G\rho_s r_s^2\mu m_H\over k}\ .
\end{equation}
Then the dimensionless temperature function $\tilde{T}(x,T_s)$ is
defined by
\begin{equation}
\label{Eqn:dimensionless temperature}
\tilde{T}(x,T_s) \equiv {T(xr_s)\over T_s}\ .
\end{equation}
If we define $\eta$ as the ratio of the total mass scale length
to the gas core radius,
\begin{equation}
\eta \equiv {r_s\over r_c}\ ,
\end{equation}
then the gas density can be written as
\begin{equation}
\label{Eqn:dimensionless gas density}
\rho_g(r) = \rho_{g0}(1+\eta^2x^2)^{-1}\ ,
\end{equation}
where we have used our assumption that $\beta=2/3$.
Using equations (\ref{Eqn:enclosed mass}) through
(\ref{Eqn:dimensionless gas density}), the dimensionless form
of the hydrostatic equation becomes
\begin{equation}
\label{Eqn:dimensionless gas hydrostatic}
{d\over dx}\left[{\tilde{T}\over 1+\eta^2x^2}\right]
	= - {(1+x)\ln(1+x) - x\over x^2(1+x)(1+\eta^2x^2)}\ .
\end{equation}
In the case of $\beta=2/3$ which we are considering here, the solution
to this equation (with the boundary conditions discussed below) can be
written in closed form in terms of Clausen's integral.
However, for the purpose of initializing our simulations it is
more straightforward to solve this equation numerically.

The outer boundary condition for equation
(\ref{Eqn:dimensionless gas hydrostatic})
depends on the pressure at large radii in the cluster.
One class of solutions would have the pressure approach
a constant value at large radii to maintain pressure equilibrium
with surrounding intercluster gas.
Another might have the pressure go to zero at a finite
radius, which would represent the outermost extent of the hot gas
in the cluster.
This condition might approximate a cluster with a strong accretion
shock at its outer radius, with the gas pressure being very small
outside of this radius.
However, the overdensity in clusters is large ($\ga 10^3$ in the
core), so the outer pressure is likely to be very much smaller
than the central pressure.
Thus, we adopt the boundary conditions that apply to an
isolated cluster that extends to large radii, so that the pressure
approaches zero at large radii.
Assuming the adopted form for the gas density profile
(eq.~[\ref{Eqn:beta model}]), this implies that
$\tilde{T}(x)\rightarrow 0$ as $x\rightarrow\infty$.
With this boundary condition,
the temperature profile requires no additional scaling parameters
beyond those supplied by the assumed density profiles, and it is
independent of the normalization of the gas profile
($\rho_{g0}$).

We constrain the allowable values of $\eta$ by noting that
convective stability requires
\begin{equation}
\eta \ga 0.71\ ,
\end{equation}
while in order for the central cooling time to be larger than
a Hubble time (to avoid producing a cooling flow), we must have
\begin{equation}
\eta \la 9.6h^{-3}
  \left({\rho_s\over10^{15}M_\odot(h^{-1}{\rm\ Mpc})^{-3}}\right)
  \left({r_s\over h^{-1}{\rm\ Mpc}}\right)^2
  \left({\rho_{g0}\over10^{15}M_\odot(h^{-1}{\rm\ Mpc})^{-3}}\right)^{-2}
\end{equation}
(Ricker \& Sarazin 2001).
By experimenting with typical values of $\rho_s$, $\rho_{g0}$, and
$r_s$, we have found $\eta=2$ to be a safe value given the resolutions
we obtain in the simulations described in \S~\ref{Sec:simulations}.
This is comparable to the sizes of the numerical core radii seen
in some large-scale structure calculations (e.g., NFW).

The resulting gas temperature profile (scaled to the central temperature)
is shown as the dotted line in
Figure \ref{Fig:single cluster plots}e (below).
For $x\la 2/\eta$ the temperature drops slowly (roughly as $x^{-0.06}$),
then flattens out and eventually tends toward an asymptotic
$\ln(x)/x$ behavior.
By the radius $x=20/\eta$, the temperature has dropped to 25\% of its
central value.

We follow a similar procedure to determine the dark matter velocity
dispersion profile.
We assume the dark matter velocity components to be isotropic and Gaussian
distributed with velocity dispersion $\sigma^2(r)$.
The virial equation for the dark matter particles is then
\begin{equation}
{d\over dr}\left[{\rho_{\rm dm}(r)\sigma^2(r)}\right] =
	-{GM(r)\over r^2}\rho_{\rm dm}(r)\ .
\end{equation}
For the dark matter the density profile we use contains $\rho_s$,
$\rho_{g0}$, $r_s$, and $r_c$; hence no simple, natural scaling
for $\sigma$ is available (as in eq.~[\ref{Eqn:temperature scaling}] below).
Therefore we choose the velocity dispersion at the cutoff radius $r=R$,
$\sigma^2_R\equiv\sigma^2(R)$, and constrain its value using the assumption
of virial equilibrium (eq.~[\ref{Eqn:virial sigma}]).
As with the gas hydrostatic equation, we numerically integrate the
resulting dimensionless virial equation.
The result (for $R/r_s=10$, which we motivate in the next section)
appears as a dotted
line in Figure \ref{Fig:single cluster plots}b.
The velocity dispersion rises from its central value to about
twice this value at $x\approx 1$, then drops slowly back to
its central value at $r=R$.
The difference in behavior at small radii in comparison to the gas
temperature is due to the differing behaviors of the dark matter and
gas densities near the center.
The dark matter requires a `colder' center to maintain its cuspy
density profile, while the gas requires a hotter center in order to
maintain its constant-density core.

\subsubsection{Scaling the Cluster Model}
\label{Sec:cluster scaling}

By choosing the dark matter and gas density profiles, we introduce
as model parameters the density scales $\rho_s$ and $\rho_{g0}$, and the
length scales $r_s$ and $r_c$.
Because both density laws (for typical values of $\beta$) yield
divergent total masses for $r\rightarrow\infty$, we introduce a
cutoff in the density profile at $r=R$, making $R$ an additional
parameter. We examine the effects of this cutoff
in the single-cluster test calculations discussed in
\S~\ref{Sec:convergence}. The velocity dispersion
profile introduces the additional `thermal' parameter
$\sigma_R$, yielding a total of six parameters.
(The scaling parameter $T_s$ for the gas temperature profile is
determined by $\rho_s$ and $r_s$ via equation
[\ref{Eqn:temperature scaling}].)

While we could examine the evolution of clusters with chosen
values of the above parameters, it is more useful to select values
for the total mass $M$, X-ray luminosity $L_X$,
emission-weighted temperature $T_X$, and gas fraction $F_g$,
expressing the density, thermal,
and length scales in terms of these quantities.
This enables us more readily to compare our results with real clusters.
In addition, it permits us to use observed correlations between these
global quantities to reduce the dimensionality of the parameter
space we must study.

We begin by defining the dimensionless density and
dark matter velocity dispersion profiles:
\begin{eqnarray}
\tilde{\rho}(x)		  &\equiv& {\rho(xr_s)\over\rho_s} \\
\tilde{\rho}_g(x,f_g,\eta)&\equiv& {\rho_g(xr_s)\over\rho_s} \\
\tilde{\sigma}(x,\sigma_R)&\equiv& {\sigma(xr_s)\over\sigma_R}\ .
\end{eqnarray}
The dimensionless gas temperature profile is defined in equation
(\ref{Eqn:dimensionless temperature}).
Here we define
\begin{eqnarray}
f_g &\equiv& {\rho_{g0}\over \rho_s}\\
c   &\equiv& {R\over r_s}\ .
\end{eqnarray}
If we take $R$ to be the virial radius $r_{200}$, $c$ becomes the familiar
halo concentration parameter. While NFW find in their simulations
that $c$ has a weak dependence on halo mass, we adopt the simplification
of a constant value $c=10$, consistent with their results for
cluster-mass halos.
Together with our constraint on $\eta$
(\S~\ref{Sec:temperature profiles}), this reduces the number of free
parameters to four.

We can now write expressions for
our desired quantities $M$, $L_X$, $T_X$, and $F_g$:
\begin{eqnarray}
M   &=& 4\pi\rho_s r_s^3\int_0^c dx\,x^2\tilde{\rho}(x)\\
L_X &=& 4\pi\rho_s^2 r_s^3\int_0^c dx\,x^2\tilde{\rho}_g(x,f_g,\eta)^2
	\Lambda[T(x)] \\
T_X &=& {4\pi\rho_s^2 r_s^3\over L_X}\int_0^c dx\,x^2
	\tilde{\rho}_g(x,f_g,\eta)^2\Lambda[T(x)]T(x)\\
F_g &=& {\int_0^c dx\,x^2\tilde{\rho}_g(x,f_g,\eta) \over
	\int_0^c dx\,x^2\tilde{\rho}(x)}\ .
\end{eqnarray}
Here $\Lambda(T)$ is the plasma emissivity function;
we use the emissivity function of the MEKAL model in XSPEC
(Mewe, Kaastra, \& Liedahl 1995) with half-solar abundances.
With the chosen values of $\eta$ (\S~\ref{Sec:temperature profiles})
and $c$, only three of these equations are independent; one of
the four quantities on the left-hand side, say $T_X$, depends on the
others.
Therefore, given $M$, $L_X$, $F_g$, and our constraints on $\eta$ and
$c$, we can numerically solve the three independent equations for
$\rho_s$, $r_s$, and $f_g$. We also have $\sigma_R$ as a
remaining free parameter.

We can further reduce the number of free parameters by considering
`typical' virialized clusters. Because $M$ and $F_g$ are observationally
the most poorly determined quantities, it makes sense to allow
them to vary. We constrain $L_X$ in terms of $M$
using the observed luminosity-temperature relation and the
virial mass-temperature relation:
\begin{eqnarray}
\label{Eqn:observed L-T}
L_X &=& K_L T_X^{a_L} \\
\label{Eqn:virial M-T}
M   &=& K_M T_X^{a_M}\ .
\end{eqnarray}
The observed exponent of the $L_X-T_X$ relation lies in the range
$\sim$2.5 to $\sim$3.0 (Markevitch 1998; Arnaud \& Evrard 1999;
Allen \& Fabian 1999) when corrected for effects due to cooling flows.
We have used the bolometric relation
determined by Markevitch (1998) from a sample of 35 nearby clusters.
In our notation Markevitch's relation corresponds to
$K_L = 7.62\times10^{42}$~erg~s$^{-1}$~keV$^{-2.64}$ and $a_L = 2.64$.
Mushotzky and Scharf (1997) show the cluster $L_X-T_X$ relation to
be nearly independent of redshift up to $z\sim 0.4$; hence this is
likely to be a good approximation even for clusters at moderate
redshifts.

The virial mass-temperature relation we use is taken from
Evrard, Metzler, \& Navarro (1996), who find in an ensemble of 58
simulated clusters that equation (\ref{Eqn:virial M-T}) with
$K_M = 7.02\times10^{13}M_\odot$~keV$^{-3/2}$ and $a_M = 3/2$
predicts cluster masses to within
a standard deviation of 15\%, even for fairly irregular clusters.

Finally, we constrain $\sigma_R$,
the scaling factor for the dark matter velocity dispersion,
using the assumption of virial equilibrium within $r=R$. We also
assume that dark matter particle orbits at $r=R$ are nearly
isotropic. This yields the constraint
\begin{equation}
\label{Eqn:virial sigma}
\sigma^2_R = {GM\over 3R}\ .
\end{equation}
In setting up the initial particle distribution for each cluster,
we choose the $x$-, $y$-, and $z$-components of the velocity for
each particle from a Gaussian distribution with dispersion
$\sigma^2(r)$.


\subsection{Choice of Parameters}
\label{Sec:choice of parameters}

\subsubsection{Model Clusters}

With the assumptions discussed above, the properties of each cluster are
completely determined by the assumed mass $M$ and gas fraction $F_g$.
We chose gas fractions which fell within the range of values
permitted by X-ray cluster observations
(David, Jones, \& Forman 1995; White \& Fabian 1995) and
which yielded central cooling times $t_{\rm cool,0}$ larger
than a Hubble time, $H_0^{-1}$.

Table~\ref{Table:cluster parameters} presents the parameter values
for the two cluster models studied in the collision runs.
Any other cluster properties can be derived from these values using
the scaling relations discussed above.
The temperatures of the two clusters were chosen to fall
in the range of `typical' rich clusters seen at low redshift
(e.g., Edge et al.\ 1990; Edge, Stewart, \& Fabian 1992).
We then chose masses using the virial mass-temperature relation
as described in section \ref{Sec:cluster scaling}.
The values of the scaling radius $r_s$ and central density
(given in Table \ref{Table:cluster parameters} in terms of
the central electron number density, $n_{e0}$) were constrained
as described in \S~\ref{Sec:cluster scaling}.
Table~\ref{Table:cluster parameters} also shows the calculated
values of the sound-crossing time $t_{\rm sc}\equiv R/c_{s0}$ and
the 2--10~keV X-ray luminosity $L_X^{2-10}$.
The two clusters, which we label `A' and `B', have a mass ratio
of approximately 2.8.
Among real nearby clusters, our simulated clusters A and B are
most nearly similar to Virgo and AWM7, respectively, although
both of these real clusters have weak cooling flows.


\begin{deluxetable}{ccccccccc}
\tablewidth{0pt}
\tablecolumns{9}
\tablecaption{Scaling Parameters for Model Clusters}
\tablehead{
\colhead{} &
\colhead{$M$} &
\colhead{$T_X$} &
\colhead{$F_g$} &
\colhead{$r_s$} &
\colhead{$n_{e0}$} &
\colhead{$t_{\rm sc}$} &
\colhead{$t_{\rm cool,0}$} &
\colhead{$L_X$ (2--10 keV)} \\
\colhead{Cluster} &
\colhead{($10^{15}M_\odot$)} &
\colhead{(keV)} &
\colhead{} &
\colhead{($h^{-1}$ kpc)} &
\colhead{($10^{-3}$ cm$^{-3}$)} &
\colhead{(Gyr)} &
\colhead{(Gyr)} &
\colhead{($10^{44}$ erg s$^{-1}$)} \\
\cline{1-9}
}
\startdata
A	& 0.199	& 2.0	& 0.125	& 118	& 3.95	& 1.89	& 17.7	& 0.117 \\
B	& 0.562	& 4.0	& 0.174	& 169	& 5.31	& 1.93	& 19.7	& 1.25 \\
\enddata
\label{Table:cluster parameters}
\end{deluxetable}

\subsubsection{Initial Merger Kinematics}
\label{Sec:initial_merger}

\begin{figure}[t]
\epsscale{0.5}
\plotone{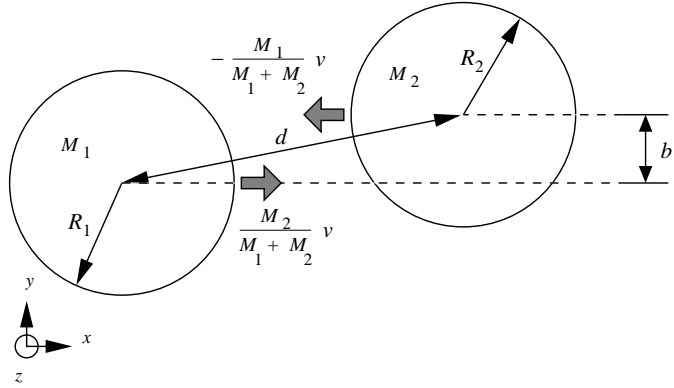}
\caption{
  Schematic of the initial conditions used for cluster merger calculations.
  The clusters have masses $M_1$ and $M_2$ and cutoff radii $R_1$ and $R_2$.
  The initial separation of the cluster centers
  is $d$, and the impact parameter ($y$-direction) is
  $b$; the collision plane is the $xy$-plane.
  The initial relative velocity is $v$; the
  simulation takes place in the center-of-mass frame.
  \label{Fig:initial conditions}
  }
\end{figure}

Our collision model, depicted in Figure \ref{Fig:initial conditions},
is appropriate for two clusters that have fallen together
from a large distance with nonzero angular momentum.
This model introduces three new parameters: the initial separation $d$,
the initial relative velocity $v$, and the impact parameter $b$.
Because clusters of given masses do not all merge at the same time
or with the same total angular momentum, we have some freedom in
choosing the values of these parameters. However, only certain ranges of
values are physically well-motivated. Here we discuss the scaling
of our choices with the properties of our model clusters.

For computational reasons, it makes sense
to fix the separation $d$ at the start of the calculation,
balancing the need to save computational time in
following the relatively uninteresting pre-merger cluster evolution
with the need to start the clusters from a virialized state.
In each case we set
\begin{equation}
\label{Eqn:initial separation}
d = \sqrt{\left(R_1+R_2\right)^2 + b^2}\ ,
\end{equation}
where $R_1$ and $R_2$ are the cutoff radii of the two clusters.
We assume the clusters to have fallen to this separation from a
greater separation $d_0 \approx 1.6d$, at which their relative radial velocity
is zero.
The value of $d_0$ corresponds to the turn-around distance for the clusters
(Sarazin 2001),
assuming that the cluster merger (the first core crossing) occurs when
the age of the Universe is $\sim$11 Gyr
(the present age for $\Omega=1$ and $h=0.6$).
The exact value of $d_0$ does not affect the total energetics much as long
as the relative velocity of the clusters at closest approach is close to the
value expected for free-fall from infinity.

For the purpose of computing the initial relative velocity,
we approximate the two clusters as point masses.
Comparing the angular momentum and energy predicted with this approximation to the
actual value determined from the simulations shows that this estimate
is good to within 1\% because of the highly concentrated nature of the
cluster density profiles.
We also neglect the clusters' spin angular momenta.
At the separation $d_0$, the clusters are assumed to have zero relative
radial velocity; hence their angular momentum and energy are
\begin{eqnarray}
\label{Eqn:init ang mom and energy}
J_0 &\approx& mv_0 d_0\\
\nonumber
E_0 &\approx& {1\over2}mv_0^2 - {GM_1M_2\over d_0}\ ,
\end{eqnarray}
where their reduced mass is
\begin{equation}
\label{Eqn:reduced mass}
m \equiv {M_1M_2\over M_1+M_2}\ ,
\end{equation}
and $v_0$ is their relative azimuthal velocity.
At the separation $d$, the relative velocity $v$ is perpendicular
to the direction of $b$, so we can write
\begin{eqnarray}
\label{Eqn:ang mom and energy}
J &\approx& mvb\\
\nonumber
E &\approx& {1\over2}mv^2 - {GM_1M_2\over d}\ .
\end{eqnarray}
Conserving angular momentum and energy, we eliminate $v_0$ and find
\begin{equation}
\label{Eqn:initial velocity}
v^2 = 2G\left(M_1+M_2\right)\left({1\over d}-{1\over d_0}\right)
      \left[1-\left(b\over d_0\right)^2\right]^{-1}\ .
\end{equation}

We estimate the range of impact parameters $b$ to study by using
the linear-theory result for the dimensionless spin of dark-matter
halos. The spin parameter $\lambda$ is defined as (Peebles 1969)
\begin{equation}
\label{Eqn:dimensionless spin}
\lambda \equiv {J|E|^{1/2}\over GM^{5/2}}\ .
\end{equation}
Here $J$ is the angular momentum of the halo, $E$ is its total
energy, and $M$ is its mass.
Recently, Sugerman, Summers, and Kamionkowski (2000) have performed a detailed
comparison of linear-theory predictions to actual angular momenta of
galaxies formed in cosmological $N$-body/hydro calculations.
These simulations did not include cooling or star formation, so at the
upper end of the mass range they studied their results should carry over
to the clusters we are simulating.
They find, in agreement with White (1984), that linear theory overpredicts
the final angular momentum of galaxies by roughly a factor of three, with
a large ($\sim 50\%$) dispersion in the ratio of the linear-theory prediction
to the actual value. The importance of nonlinear effects on the spin
parameter is less clear, as uncertainties in their group-finding algorithm
strongly affect the determination of $J$, $E$, and $M$ for each galaxy and
produce a large scatter in $\lambda$. However, given this uncertainty,
the mean value of $\lambda$ they obtain is about 0.05, close to the
linear-theory result.

We take the halo to be the final merger remnant.
Its final total angular momentum is the sum of the spin angular momenta of
the two subclusters plus the orbital angular momentum $J_{\rm orb}$.
We estimate the initial spins of the subcluster by applying
equation~(\ref{Eqn:dimensionless spin}) to each of them, and we assume
that the initial spins are correlated.
The orbital angular momentum is then the difference between the
final angular momentum of the merger remnant, and the total spin angular 
momenta of the subclusters
(Sarazin 2001).
Substituting equations
(\ref{Eqn:init ang mom and energy})-(\ref{Eqn:initial velocity})
into equation (\ref{Eqn:dimensionless spin})
and solving for the impact parameter, we find that
\begin{equation}
\label{Eqn:impact parameter}
b \approx \lambda \,
\left(\frac{d_0 d}{2} \right)^{1/2} \,
\left( 1 - \frac{d}{d_0} \right)^{-1/2} \,
f(M_1,M_2) \, .
\end{equation}
Here, the function $f(M_1, M_2 )$ corrects for the internal angular momenta
and energy of the subclusters, and can be written as
\begin{equation}
\label{Eqn:fcorr}
f( M_1 , M_2 ) \equiv \frac{ ( M_1 + M_2 )^3}{ M_1^{3/2} M_2^{3/2}} \,
\left[ 1 -
\frac{ \left( M_1^{5/3} + M_2^{5/3} \right)}{\left( M_1 + M_2
\right)^{5/3}}
\right]^{3/2} \, ,
\end{equation}
but it only depends on the ratio $( M_< / M_> )$ of the smaller to
larger mass of the two subclusters.
It varies between
$4( 2^{2/3} - 1 )^{3/2} \approx 1.80 \le f(M_1, M_2) \le (5/3)^{3/2}
\approx 2.15$,
so that $f( M_1, M_2 ) \approx 2$.
In our merger models, $d \approx 20 r_{s,{\rm max}}$ ($r_{s,{\rm max}}$
is the larger of the two scale radii of the clusters), and $d_0 \approx 1.6 d$.
If we assume $\lambda \approx 0.05$ and approximate
$f( M_1, M_2 ) \approx 2$, we find that the typical impact parameter is
$b \approx 3 r_{s,{\rm max}}$.
Since this argument is statistical and a range of values are expected
in mergers, we consider values of $b = (0,2,5)\, r_{s,{\rm max}}$
in our simulations.


\begin{deluxetable}{cccccccccc}
\tablewidth{0pt}
\tablecolumns{9}
\tablecaption{Cluster Parameters Used for Collision Runs}
\tablehead{
\colhead{} &
\colhead{Cluster} &
\colhead{Mass} &
\colhead{} &
\colhead{$v$} &
\colhead{$\ell_x$} &
\colhead{$\ell_y$} &
\colhead{$\ell_z$} &
\colhead{} \\
\colhead{Identifier} &
\colhead{Models} &
\colhead{Ratio} &
\colhead{$b/r_{s,{\rm max}}$$^{\rm a}$} &
\colhead{($h^{-1}$ Mpc Gyr$^{-1}$)} &
\multicolumn{3}{c}{($h^{-1}$ Mpc)$^{\rm b}$} &
\colhead{$\Delta_{\rm min}/r_{s,{\rm min}}$$^{\rm a,c}$} \\
\cline{1-9}
}
\startdata
C1 & AA	& $1:1$	& 0 & 0.436 & 7.10 & 4.73 & 4.73 & 0.144 \\
C2 & AA	& $1:1$	& 2 & 0.436 & 7.10 & 4.97 & 4.73 & 0.144 \\
C3 & AA	& $1:1$	& 5 & 0.436 & 7.10 & 5.32 & 4.73 & 0.144 \\
C4 & AB	& $1:3$	& 0 & 0.548 & 8.63 & 6.78 & 6.78 & 0.213 \\
C5 & AB	& $1:3$	& 2 & 0.548 & 8.63 & 7.12 & 6.78 & 0.213 \\
C6 & AB	& $1:3$	& 5 & 0.548 & 8.63 & 7.63 & 6.78 & 0.213 \\
\enddata
\\
\tablenotetext{a}{$r_{s,{\rm max}}$ and $r_{s,{\rm min}}$ are the larger
  and smaller of the two scale radii, respectively.}
\tablenotetext{b}{$\ell_x$, $\ell_y$, and $\ell_z$ are the physical dimensions
  of the computational grid.}
\tablenotetext{c}{$\Delta_{\rm min}$ is the smallest zone spacing in any
  direction.}
\label{Table:collision parameters}
\end{deluxetable}
\noindent

Table~\ref{Table:collision parameters} presents the parameter values used
in the collision runs.
The simulations are labeled `C1' to `C6'; runs C1--C3 are equal-mass
collisions using cluster A, while runs C4--C6 use clusters A and B.
Each run is followed for approximately 15~Gyr, more than the age of
the Universe for $\Omega=1$ and $h=0.6$.
For the hydrodynamics and gravitational potential calculations,
we use a nonuniform mesh with $256\times128^2$ zones covering a volume
measuring approximately $6R_{\rm max}\times\left(4R_{\rm max}\right)^2$,
where $R_{\rm max}$ is the larger of $R_1$ and $R_2$.
The minimum zone spacing is approximately equal to one-seventh of the
larger of the two dark matter scale radii, or $\sim 17h^{-1}$~kpc for
runs C1--C3 and $\sim 25h^{-1}$~kpc for runs C4--C6.
For the dark matter, each run also uses $128^3$ particles of equal mass,
distributed between the merging clusters in proportion to their total
dark mass.
The results of these runs are described in \S\S~\ref{Sec:equal-mass}
and \ref{Sec:unequal-mass}.
In \S~\ref{Sec:convergence} we first discuss the results of a convergence
study involving a single cluster, which we used to set the grid and particle
parameters used in the collision runs.


\section{Numerical Simulations}
\label{Sec:simulations}


\subsection{Single-Cluster Convergence Study}
\label{Sec:convergence}

We carried out several calculations of the motion of a single cluster
with different spatial resolutions and numbers of particle in order to
evaluate numerical effects due to our boundary conditions and finite-radius
cutoff.
These runs also enabled us to choose grid and particle parameters for the
collision calculations.
These runs used the cluster parameters appropriate to cluster A (see
\S~\ref{Sec:choice of parameters}) together with the numerical
parameters listed in Table \ref{Table:single-cluster grids}.
Six runs were performed with two different grid spacings and three different
numbers of particles. One additional run was performed at a grid spacing
and particle number slightly better than the collision runs to be described
later. 
The computational volume measured $7.08\times4.72\times4.72\,
(h^{-1}\,{\rm\ Mpc})^3$, and the cluster started at the position
(2.36,2.36,2.36)$\,h^{-1}\,{\rm Mpc}$ with an initial velocity of
0.436$\,h^{-1}\,{\rm Mpc\ Gyr}^{-1}$ in the $x$-direction.
In each run the cluster was permitted to move for 5~Gyr, yielding
an expected final displacement of 2.18$\,h^{-1}\,{\rm Mpc}$.


\begin{deluxetable}{crrr}
\tablewidth{0pt}
\tablecolumns{4}
\tablecaption{Mesh/Particle Parameters Used for Single-Cluster Runs}
\tablehead{
\colhead{Identifier} &
\colhead{Mesh Size} &
\colhead{No.\ of Particles} &
\colhead{$\Delta_{\rm min}/r_s$$^{\rm a}$} \\
\cline{1-4}
}
\startdata
S1	& $64\times32\times32$		& 65536		& 0.80 \\
S2	& $64\times32\times32$		& 524288	& 0.80 \\
S3	& $64\times32\times32$		& 1048576	& 0.80 \\
S4	& $128\times64\times64$		& 65536		& 0.25 \\
S5	& $128\times64\times64$		& 524288	& 0.25 \\
S6	& $128\times64\times64$		& 1048576	& 0.25 \\
S7	& $256\times128\times128$	& 1048576	& 0.12 \\
\enddata
\tablenotetext{a}{$\Delta_{\rm min}$ is the smallest zone spacing in any
  direction.}
\label{Table:single-cluster grids}
\end{deluxetable}

\begin{figure}[t]
\epsscale{0.75}
\plotone{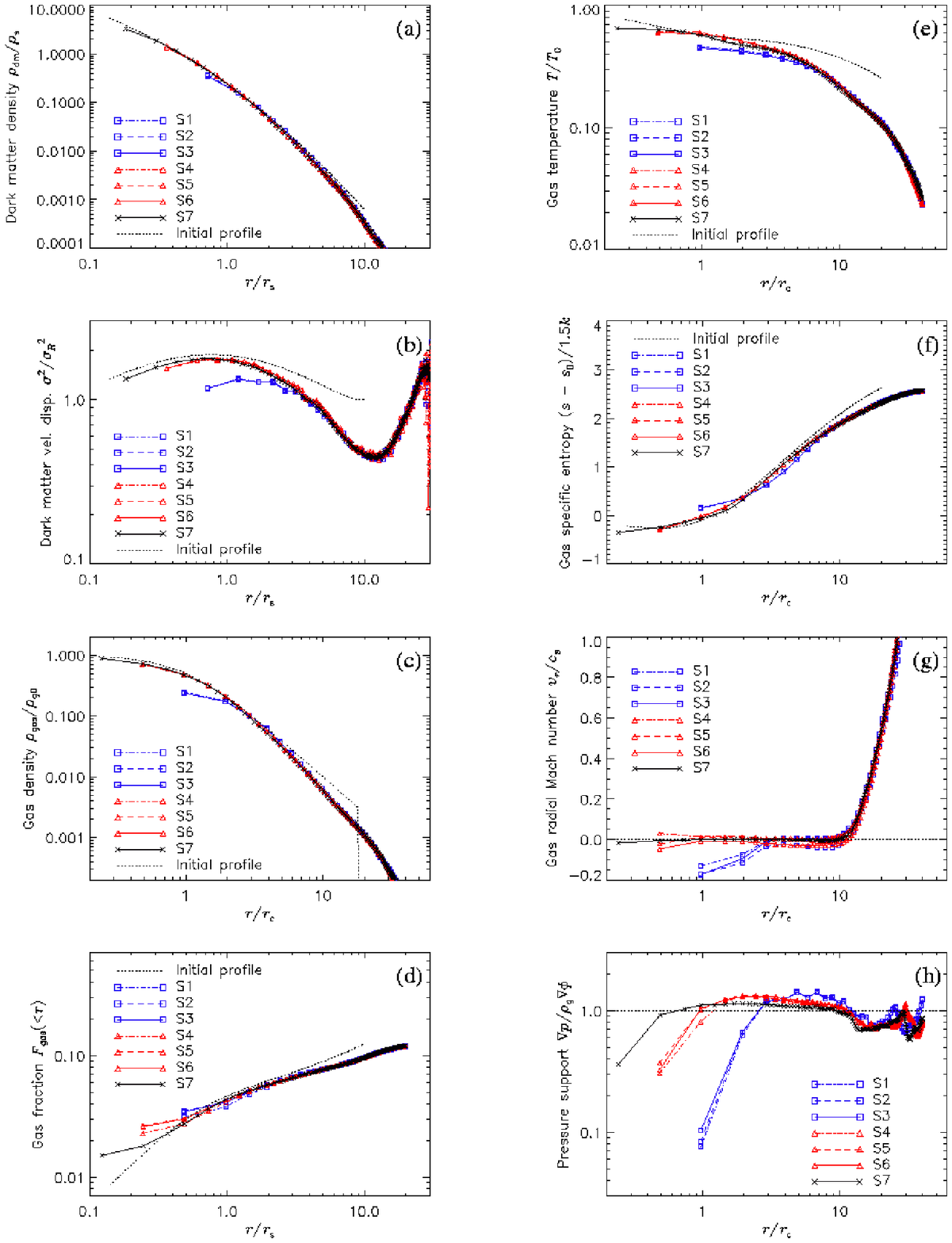}
\caption{
  Average profiles of dark matter and gas quantities at $t=5$~Gyr in the
  single-cluster test runs.
  Parameters used in the test runs are given in Table
  \ref{Table:single-cluster grids}. In all cases except (d), (g), and (h),
  quantities are scaled to their initial characteristic values.
  (a) Dark matter density. (b) Dark matter velocity dispersion.
  (c) Gas density. (d) Enclosed gas fraction. (e) Gas temperature.
  (f) Gas specific entropy. (g) Gas radial Mach number.
  (h) Ratio of pressure force to gravitational force on the gas.
  \label{Fig:single cluster plots}
  }
\end{figure}

Figure \ref{Fig:single cluster plots} compares final average profiles
of various quantities in the single-cluster tests to the initial profiles.
Included are the dark matter density and velocity dispersion, gas density,
enclosed gas fraction, gas temperature and specific entropy, and the gas
radial Mach number. We also plot the ratio of the gas pressure gradient
to the gravitational force per unit volume on the gas, which shows the
extent to which the gas is in hydrostatic equilibrium.
The results show the effects of finite zone spacing and particle number
on small scales and outflow boundary conditions on large scales.
For gridded (gas) quantities, each profile was generated
by averaging profiles interpolated along 1000 different directions from the
gas centroid.
The radial sample spacing was
taken to be equal to the size of the smallest zones in the calculation.
For particle (dark matter) quantities, particles were binned in radial bins
about the dark matter centroid; then the particle count or the average value
of the velocity dispersion in each bin was used. The bin spacing used was
the same as the sample spacing used for the gas profiles. With the
exception of the gas fraction, the pressure-potential ratio, and the
radial Mach number, all quantities are scaled in terms of the initial
values of the appropriate scaling parameters, such as the core radius or
central gas density.

The dark matter density profile (Figure \ref{Fig:single cluster plots}a)
shows little sensitivity to the number of particles or zone spacing
for the ranges of these parameters considered here. Resolution effects
appear in the innermost part of the cluster; the density in the innermost
radial bin tends to be some 15\% below the initial value. This demonstrates
the tendency of particle-mesh force smoothing to produce an artificial
constant-density core on scales smaller than twice the mesh spacing.
Thus in order to observe the turnover to $r^{-1}$ behavior at small scales
characteristic of the NFW profile, the mesh spacing should be no larger
than about 1/4 the scale radius $r_s$.
For $r\ga 2r_s$, the escape of particles through the outer boundary of
the cluster makes the resulting final profile slightly steeper than its initial
$r^{-3}$ shape. Runs S1-S3 display a somewhat shallower profile between
$2r_s$ and $5r_s$ than that in runs S4-S7.

The dark matter velocity dispersion profile
(Figure \ref{Fig:single cluster plots}b) also reflects this evaporative
cooling: within $r\approx r_s$ in run S7, $\sigma^2$ is
about 9\% lower than its initial value, with roughly the same $r$ dependence.
Outside $r\approx r_s$ the final profile drops off more steeply, whereas
outside $r=R$ it appears to rise again due to the radial motion of escaping
particles. (We estimate $\sigma^2(r)$ by computing
$\langle v_i^2\rangle - \langle v_i\rangle^2$ for
each velocity component $i$, averaging within radial shells, then sum the results for
$x$, $y$, and $z$ and divide by 3. This procedure assumes isotropic particle
velocity dispersions and does not subtract out radial expansion or contraction.)
The central dispersion in runs S1-S3 is substantially lower than in run S7,
but runs S4-S6 agree with S7. 

The lack of an external confining pressure produces expansion of the
gas at the outer boundary of the cluster.
This has a major effect on the gas density profile (Figure
\ref{Fig:single cluster plots}c), causing it to become steeper at large radii.
The profile also becomes slightly shallower at small radii.
However, unlike the dark matter profile, once the gas core radius
is resolved (runs S4-S7), the gas profile converges to a central density
slightly below the initial value. Between $r\sim r_c$ and $r\sim4r_c$
it agrees with the initial profile, but the effective core radius
(defined as that radius at which the density drops to one-half its
central value) at
the end of the simulation is $\sim2\%$ smaller than the initial value,
so the overall shape at small radii differs slightly from a $\beta$-model.
When the core is not resolved (runs S1-S3), the behavior is more like the
dark matter density, with the innermost density point well below its initial
value. Again, the number of particles makes little difference for the
range of values considered here.

The different behaviors of the gas and dark matter are apparent from the
enclosed gas fraction profile (Figure \ref{Fig:single cluster plots}d).
The enclosed gas fraction $F_{\rm gas}(<r)$ is computed from the average
gas and dark matter densities via the relation
\begin{equation}
F_{\rm gas}(<r) = {\int_0^r r^2dr\,\rho_{\rm gas}(r)\over
	\int_0^r r^2dr\,\rho_{\rm tot}(r)}\ .
\end{equation}
In all cases the gas fraction at a given radius is smaller than its initial
value for radii larger than about two zones. Within two zones of the center
the gas fraction flattens out to a constant value rather than dropping to
zero because of the artificial core in the dark matter profile.
The final gas fraction rises more slowly than its initial profile, reaching
the initial total value of 12.5\% only at $r\sim2R$.

The spreading of the gas due to the lack of external pressure confinement
leads to a steepening of the temperature profile
(Figure \ref{Fig:single cluster plots}e) and the advection of entropy
out of the outer parts of the cluster (Figure \ref{Fig:single cluster plots}f).
Whereas the initial temperature profile drops to $\sim0.4$ times its central
value at $r=R$, by the end of the single-cluster calculation it has dropped
to about 0.2 times its central value. In runs S1-S3 the gas core is not
resolved, and the innermost temperature point is about two-thirds its
initial value at that radius. For runs S4-S7 the final temperature agrees
much better with its initial profile, although run S7 does not reproduce
the upturn within $\sim0.5r_c$ present in the initial conditions. The
two sets of core-resolved runs also disagree slightly between $r\sim r_c$
and $r\sim3r_c$, with runs S4-S6 agreeing better with the initial profile.
All of the simulated profiles agree with each other beyond about five
core radii. Similar results can be seen in the specific entropy profile.
Note that the specific entropy is measured relative to the central value
given the imposed density and temperature profiles. The resulting imposed
entropy profile passes through a minimum near $r=0.5r_c$, so at smaller
radii the gas would be convectively unstable if our calculations were to
use smaller zones.

The radial Mach number (Figure \ref{Fig:single cluster plots}g) and
pressure support profiles (Figure \ref{Fig:single cluster plots}h) show
that the innermost parts of the cluster approximate hydrostatic equilibrium
much better as grid resolution improves. Both quantities are formed from
angle-averaged values; thus the radial Mach number is computed using the
average radial velocity and the average temperature.
When the core is not resolved,
the innermost radial velocity point is as much as 0.2 times the sound speed,
and the pressure provides about 10\% of the gravitational force there.
In runs S4-S7 the maximum Mach number inside $r=10r_c$ is less than 5\%,
and at the innermost point the pressure contributes 30\% of the force due
to gravity. The pressure support ratio increases immediately to a value
slightly larger than unity outside the center, indicating that the outer
parts of the cluster are nearly in hydrostatic equilibrium but are still
slowly expanding.
In all cases the gas outside $r=10r_c$ is in nearly free expansion,
with the radial velocity increasing to a sonic point at $r\sim25r_c$.
The material inside this sonic point is slowly decelerating and will
eventually fall back onto the cluster.

In summary, the single-cluster test calculations show that the values of
particle number and zone spacing used in the collision runs
are adequate to produce a control cluster
which remains in nearly hydrostatic equilibrium.
The code is especially
well-converged with respect to particle number for the value ($\sim 10^6$)
used in the collision runs, and the quantities of interest have
converged down to radii of about two zones for the zone spacing used
($\sim r_c/3.5$).
The accuracy of the converged solution is affected by the particle-mesh
smoothing length on small scales and by the lack of an external confining
pressure on large scales.
These effects lead to an artificial core radius of 1--2 zones in the
dark matter and a nearly free expansion of the gas and dark matter outside
the initial density cutoff which steepens the density and temperature
profiles near the cutoff.


\subsection{Equal-Mass Mergers}
\label{Sec:equal-mass}

\begin{figure}
\epsscale{0.8}
\plotone{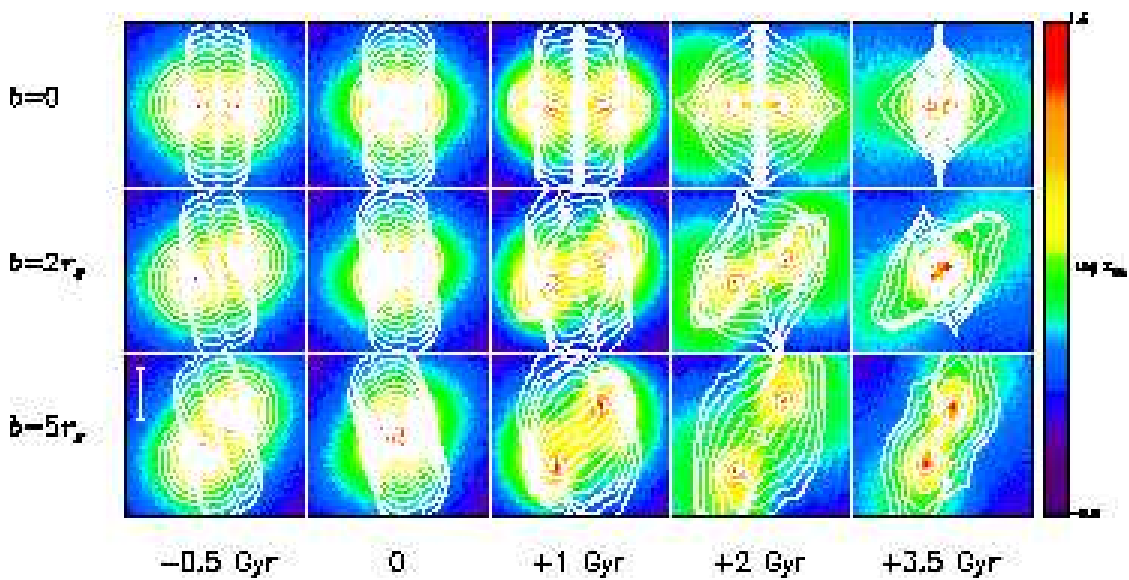}
\caption{
  Surface density maps for dark matter (colormap) and gas (contours) in the 1:1
  collision runs (C1--C3), projected along the $z$-axis.
  A small region [$\sim (h^{-1}{\rm\ Mpc})^2$] of the projected simulation
  volume is shown.
  Logarithmic color scaling for the dark matter surface density
  $\Sigma_{\rm dm}$ is indicated by the key to the right of the figure;
  units are $10^{15}\,M_\odot\,(h^{-1}{\rm Mpc})^{-2}$.
  Contours of gas surface density $\Sigma_{\rm gas}$ are spaced by a
  factor of two, with the outermost contour having the value
  $10^{-5}$ in the same units.
  The fiducial bar in the lower left corner is $0.5h^{-1}\,{\rm Mpc}$
  long.
  For each run, $t=0$ corresponds to the time at which the luminosity of
  the system reaches its maximum value.
  \label{Fig:1:1 mass plot}
  }
\end{figure}

\begin{figure}
\epsscale{0.8}
\plotone{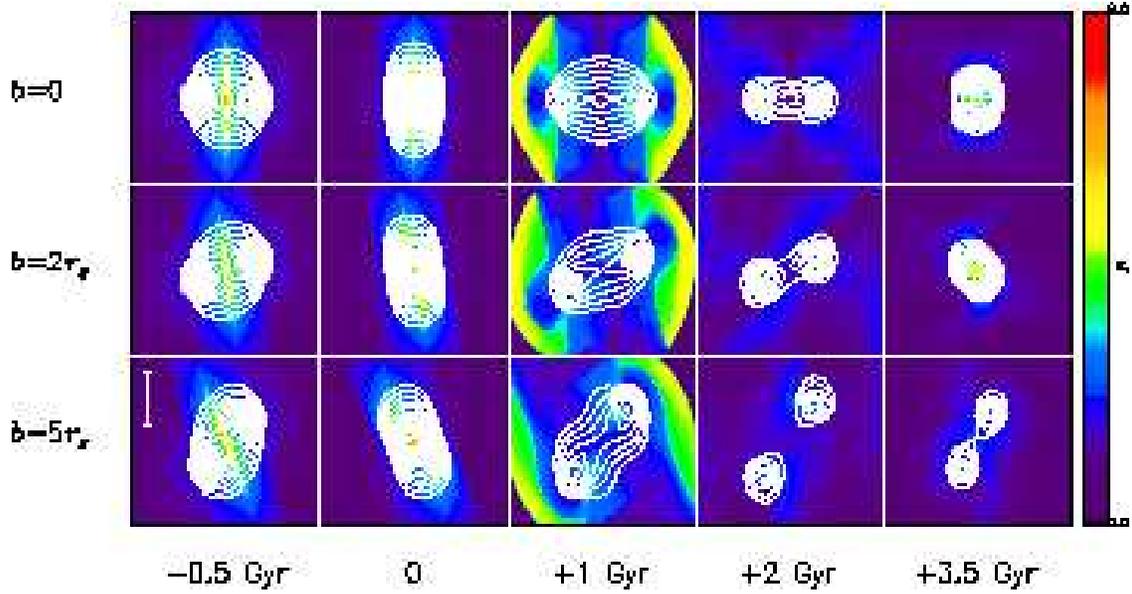}
\caption{
  2--10~keV emission-weighted temperature $T_X$ (colormap) and X-ray
  surface brightness $S_X$ (contours) in the 1:1 collision runs (C1--C3),
  as seen along the $z$-axis.
  A small region ($\sim (h^{-1}{\rm\ Mpc})^2$) of the projected simulation
  volume is shown.
  Linear color scaling for $T_X$ is indicated by the key to the right of
  the figure; units are keV.
  Surface brightness contours are spaced by a factor of three,
  with the outermost contour having the value
  $6.3\times10^{-17}$~erg~s$^{-1}$~cm$^{-2}$.
  The fiducial bar in the lower left corner is $0.5h^{-1}\,{\rm Mpc}$
  long.
  For each run, $t=0$ corresponds to the time at which the luminosity of
  the system reaches its maximum value.
  \label{Fig:1:1 xray plot}
  }
\end{figure}

\begin{figure}
\epsscale{0.5}
\plotone{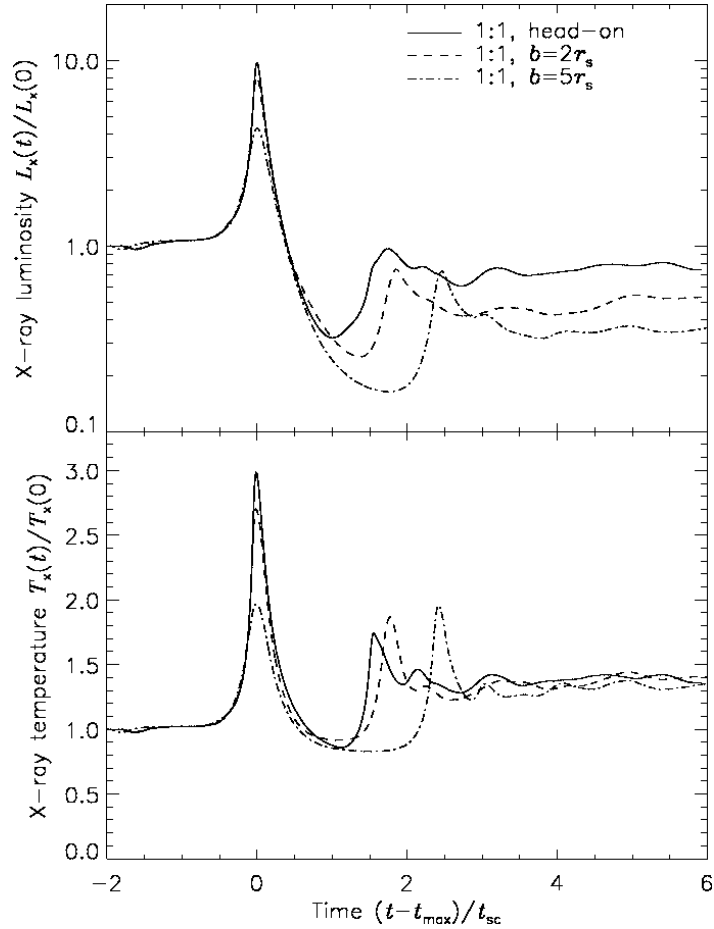}
\caption{
  2--10~keV luminosity $L_X$ and average emission-weighted temperature
  $T_X$ as functions of time $t$ in the 1:1 collision runs (C1--C3).
  Luminosity and temperature are each scaled to their initial values.
  Time is offset from the time of peak luminosity $t_{\rm max}$ and
  scaled to the sound-crossing time $t_{\rm sc}$.
  \label{Fig:1:1 lt plot}
  }
\end{figure}

In this section we describe the simulations of mergers between clusters
of equal mass (runs C1--C3).
Figure \ref{Fig:1:1 mass plot} shows as functions of time
the projection of the gas and dark
matter densities along the $z$-axis, which is perpendicular to the plane
of the collision in the offset runs.
We refer to the projected densities as $\Sigma_{\rm g}$ and
$\Sigma_{\rm dm}$, respectively.
Figure \ref{Fig:1:1 xray plot} shows the 2--10~keV X-ray surface brightness
$S_X$ and X-ray emission-weighted temperature $T_X$, also projected along
the $z$-axis.
Figure \ref{Fig:1:1 lt plot} shows the total luminosity $L_X$ and average
emission-weighted temperature for the entire computational grid as functions
of time.
The zero of time in each plot has been chosen to be the instant of maximum
luminosity (in each case approximately 3.5~Gyr or $1.8t_{\rm sc}$ after the
start of the calculation).
The abscissa of the luminosity and temperature plots is also scaled to the
sound-crossing time $t_{\rm sc}$.

As the outer regions of the two clusters begin to interact early in the
merger, a pair of roughly planar shocks forms at an angle
to the collision axis, moving in opposite directions.
By 0.5~Gyr prior to maximum luminosity, these planar shocks have passed
through the cluster cores and are on their way out of the interaction
region.
They are readily apparent in the projected gas mass plot (Figure
\ref{Fig:1:1 mass plot}) and are
barely visible in the surface brightness contours (Figure
\ref{Fig:1:1 xray plot}).
Between the clusters the temperature is higher
than the pre-shock value by about a factor of three, and the material in
this region is expelled from the collision axis at about
1/2 the initial relative velocity.
The material between the clusters is adiabatically compressed
after the passage of the shocks.
In previous gas-only merger
calculations (Ricker 1998), the planar shocks later developed
into spiral shocks which acted to dissipate the angular momentum of the
cluster cores into large-scale rotational motion.
With the addition of collisionless dark matter, however, the planar
shocks play less of a role in the later stages of the collision, and
we see more small-scale rotation
as early as 3--4~Gyr after maximum brightness and continuing for at least
3~Gyr beyond that.

Although we have not attempted to reproduce any specific clusters
with these calculations, we note that the simulations do resemble
some merging clusters with particularly simple geometries.
The temperatures and X-ray surface brightnesses in the early phases
of the zero-offset, equal mass mergers strongly resemble the
observed {\it ROSAT} image and {\it ASCA} temperature map of the cluster
surrounding Cygnus-A (Markevitch et al.\ 1999). The offset 1:1
cases, seen just after first core passage, resemble the structure
of Abell~3395 (Markevitch et al.\ 1998), which appears to be observed
from above the merger plane.

The merging clusters undergo a dramatic increase in luminosity and
temperature as the cluster cores collide.
The luminosity jump ranges from a factor of 4 for the most off-center case
to a factor of 10 for the head-on case.
The average temperature increases by a factor of two to three.
In the projected mass and X-ray maps, the period of core collision
corresponds to the
formation of a strong, crudely ellipsoidal shock which dissipates some of the
initial kinetic energy of the cores into thermal energy in the surrounding
gas.
The greater increase in total luminosity and average temperature for the
head-on case results from the fact that the dark matter cores, whose gas
content produces the majority of the X-ray emission, pass directly through
one another and the adiabatically heated gas between them.
In the offset cases this hot gas is driven ahead of the cluster cores
as they slip past one another, producing an S-shaped region of high
temperature.
In all cases, however, the dark matter cores continue to have associated
peaks in the gas density even after they pass
one another, a significant departure from the behavior in the gas-only
collisions considered by Ricker (1998).
Here the central potential is dominated by collisionless dark matter,
which dissipates energy in a relatively inefficient manner compared
to the gas during core passage.
Thus the dark matter cores, which are able to pass through one another even
in the head-on case, drag gas density peaks along with them.
The extra oscillation due to the dark matter allows the dark matter to
`pump' the gas, keeping the system disrupted longer than would be expected
if both matter components were collisional.
Offsets on the order of a few hundred kpc are sometimes observed between the
dark matter and gas density maxima, particularly as the cluster cores
reach maximum separation after their first interaction.

The first period of enhanced luminosity and temperature lasts for approximately
one sound-crossing time.
After their close interaction, the cores recede from one another,
causing the temperature to drop to about 80\% of its initial value.
By the time the cores reach their maximum separation,
the luminosity decreases much more, reaching about 15\% of its initial
value in the most offset run and about 30\% in the head-on run.
As one might expect due to the relative efficiency of gas dissipation
in the three cases, the clusters require longer to reach maximum separation
in run C3; this occurs about $t_{\rm sc}$ after core passage
in this run, compared to only 1/2 sound-crossing time for run C1.
In each case, a second core passage occurs between 1.5 and 2.5 sound-crossing
times after the first, producing smaller brightness and temperature jumps.
Following this second passage, the total luminosity drops to between
30 and 70 percent of its initial value, while the average temperature
drops to 1.4 times its initial value.
The cores continue to oscillate, producing minor variations in the luminosity
and temperature, but the values of these quantities change very little for
the remainder of the calculation.
During this time several weaker sets of shocks are generated by core passages,
as is apparent at 3.5~Gyr after the first core passage in the projected
mass and X-ray plots (Figures \ref{Fig:1:1 mass plot} and
\ref{Fig:1:1 xray plot}). We discuss the evolution of luminosity and
temperature in more detail in \S~\ref{Sec:luminosity-temperature}.

After the cores have coalesced,
angular momentum continues to be redistributed in the gas by
turbulent eddies which are fed by material falling back
in the wake of the ellipsoidal shock.
The remnant requires several billion years to reach equilibrium, and
indeed has not quite reached equilibrium by $t=15{\rm\ Gyr}$, the end
of the calculation (and more than a Hubble time for $h=0.6$).
We consider the properties of the merger remnant at $t=15{\rm\ Gyr}$
in \S~\ref{Sec:merger remnants}.


\subsection{Mergers with a 1:3 Mass Ratio}
\label{Sec:unequal-mass}

\begin{figure}
\epsscale{0.8}
\plotone{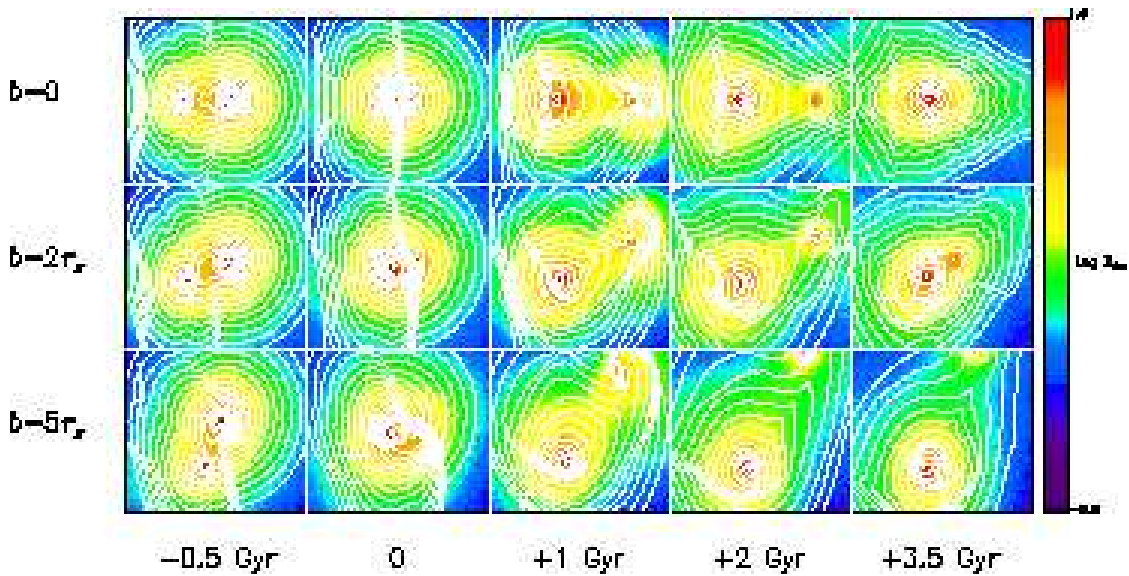}
\caption{
  Surface density maps for dark matter (colormap) and gas (contours) in the 1:3
  collision runs (C4--C6), projected along the $z$-axis.
  Scaling of the plot is the same as in Figure~\protect\ref{Fig:1:1 mass plot}.
  \label{Fig:1:3 mass plot}
  }
\end{figure}

\begin{figure}
\epsscale{0.8}
\plotone{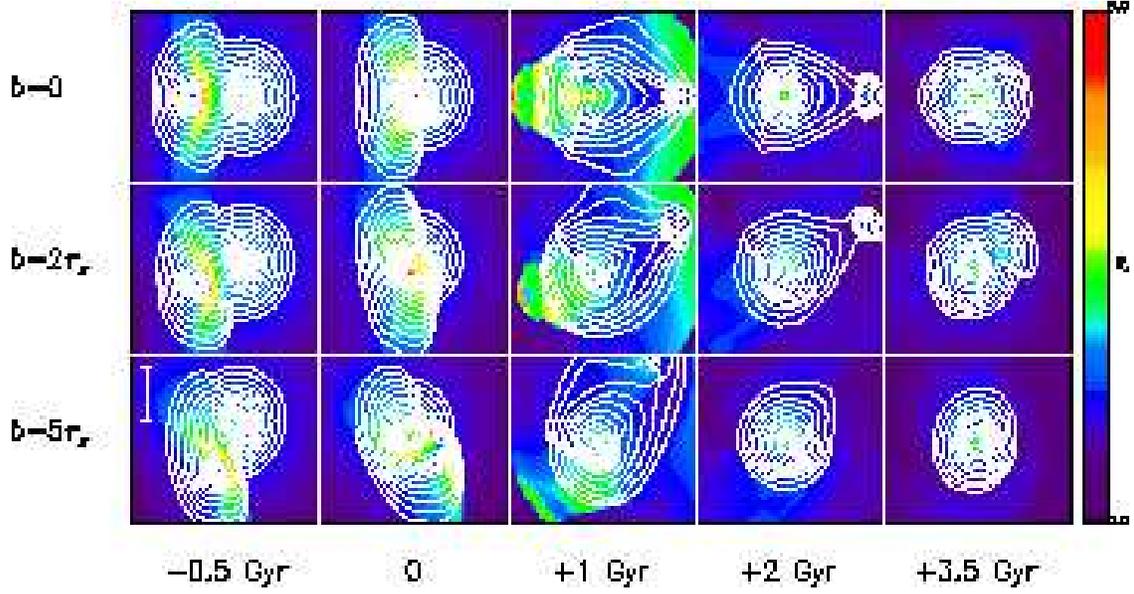}
\caption{
  2--10~keV emission-weighted temperature $T_X$ (colormap) and X-ray
  surface brightness $S_X$ (contours) in the 1:3 collision runs (C4--C6),
  as seen along the $z$-axis.
  Scaling of the plot is the same as in Figure~\protect\ref{Fig:1:1 xray plot}.
  \label{Fig:1:3 xray plot}
  }
\end{figure}

\begin{figure}
\epsscale{0.5}
\plotone{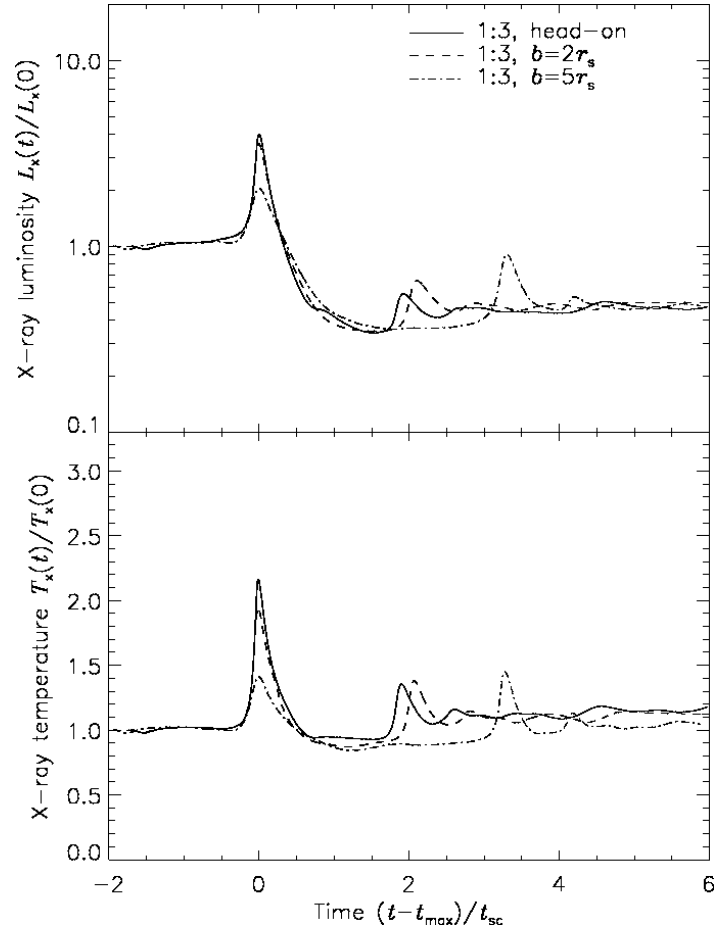}
\caption{
  2--10~keV luminosity $L_X$ and average emission-weighted temperature
  $T_X$ as functions of time $t$ in the 1:3 collision runs (C4--C6).
  Scaling of the plot is the same as in Figure
  \ref{Fig:1:1 lt plot}.
  \label{Fig:1:3 lt plot}
  }
\end{figure}

In this section we describe the simulations of mergers between clusters
with a mass ratio of approximately 1:3 (runs C4--C6).
Figure \ref{Fig:1:3 mass plot} shows mass maps projected along the $z$-axis
($\Sigma_{\rm g}$ and
$\Sigma_{\rm dm}$) for these runs at several different times.
Figure \ref{Fig:1:3 xray plot} shows the 2--10~keV X-ray surface brightness
$S_X$ and X-ray emission-weighted temperature $T_X$, also projected along
the $z$-axis.
Figure \ref{Fig:1:3 lt plot} shows the total luminosity $L_X$ and average
emission-weighted temperature as functions of time.
In these figures the more massive cluster initially approaches from the
right.
As for the 1:1 merger runs,
the zero of time in each plot has been chosen to be the instant of maximum
luminosity.
The abscissa of the luminosity and temperature plots is also scaled to the
sound-crossing time $t_{\rm sc}$ of the more massive cluster.
Numerically, the values of the
time of maximum luminosity and the sound-crossing time are
similar to those obtained in the 1:1 collisions ($\sim 3.5{\rm\ Gyr}$ and
1.9~Gyr, respectively).

In the 1:3 runs the initial pair of shocks is not planar; instead,
the shock sweeping through the less massive cluster curves around its
core, and the adiabatically compressed region between the clusters
takes on an arc shape.
For nonzero impact parameters,
the shock being driven into the more massive cluster is stronger on
the side from which the less massive cluster approaches.
However, at 0.5~Gyr prior to maximum luminosity the right-hand shock has
not yet penetrated the more massive cluster core, as it has by this point in
the equal-mass collisions.

Peak luminosity and temperature occur as the core of the less massive
cluster passes through (in the head-on case) or swings past (in the
offset cases) that of the more massive cluster.
As in the equal-mass runs, in the offset cases here an S-shaped region
of hot gas is apparent in the region between the cluster cores.
The heating of the gas is more severe in the region in front of the
less massive cluster.
Also, the less massive cluster undergoes extreme distortions at this time.
Gas velocities of up to 2000~km~s$^{-1}$ are present behind it,
and as it plows into the hotter, denser ICM of the
larger cluster, it is flattened perpendicular to its direction of motion.
The core of the larger cluster is less affected by the interaction, as
expected.
Shortly after first core passage, the offset case C5 resembles
Abell~754, consistent with the results of Roettiger et al.\ (1998).
The offset cases also resemble Abell~2256 just at core passage if
one uses the {\it ASCA} or {\it BeppoSAX} temperature maps (Markevitch 1996;
Molendi, de Grandi, \&
Fusco-Femiano 2000) and assumes the merger axis to be perpendicular to
the plane of the sky.

Although the smaller cluster undergoes severe distortion during the
core passage, the merger's effect on the total luminosity and
average emission-weighted
temperature of the system is noticeably less than in the 1:1 collisions.
The luminosity increases by a factor of 2--4, and the temperature by a
factor of 1.4--2.2, for about 1/2 sound-crossing time during the initial
core passage.
While maximum separation following core passage is reached at about the
same time as in the 1:1 cases, the drop in luminosity and temperature
is not as great.
This period is again followed by a weaker secondary peak in luminosity
and temperature as the cores fall back together.
Here the most offset case (run C6) spends more time at maximum separation,
causing the spread in secondary peak times to be 50\% greater for the
1:3 runs than in the 1:1 cases.
The second core passage for this run takes place almost 7~Gyr after the
first, and in all cases this second interaction takes place on a nearly
radial trajectory due to the dissipation caused by the first interaction.
(The last two panels of Figures \ref{Fig:1:3 mass plot} and
\ref{Fig:1:3 xray plot} for this run do
not show the less massive cluster because at these times it is at a
separation which places it outside the plotted region.)
Both luminosity and temperature settle down to final values of
50\% and 110\%, respectively, of their initial values following the second
core passage.
A few oscillations are again seen as the cores settle into a common remnant.
Unlike the 1:1 cases, here there is little variation in the final
luminosity as impact parameter is varied.


\section{Discussion}
\label{Sec:discussion}


\subsection{Structure of Merger Remnants}
\label{Sec:merger remnants}

\begin{figure}[t]
\epsscale{0.7}
\plotone{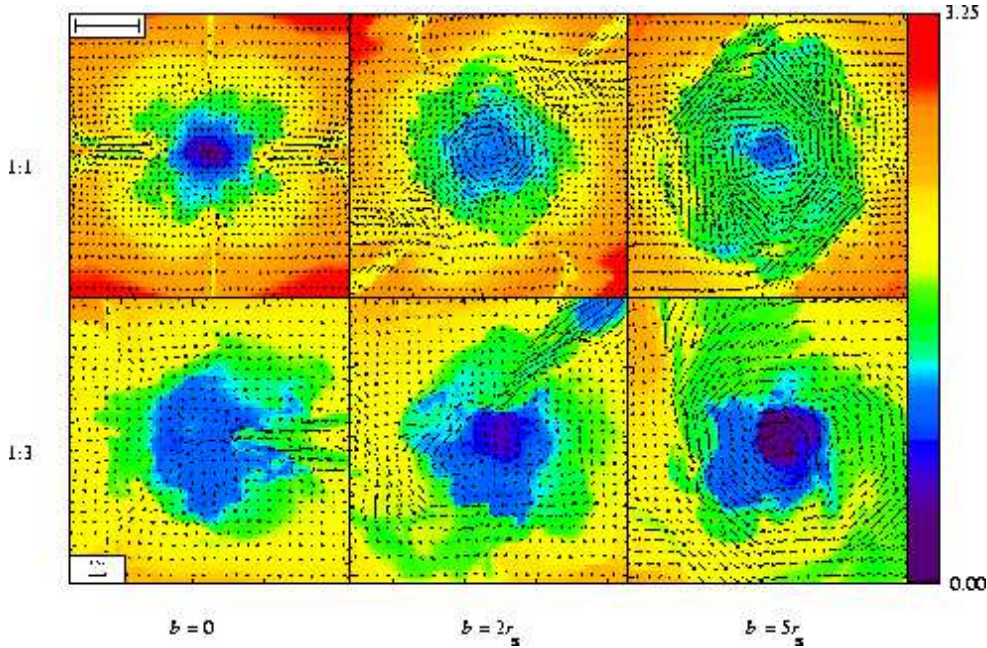}
\caption{
  Gas specific entropy and velocity fields in the collision plane ($xy$)
  of each run at $t=14.5$~Gyr, near the end of the simulations.
  Entropy (colormap) is measured relative to the initial central value of
  the more massive cluster in each run in units of $1.5k$.
  The color key to the right of the figure indicates the range of values
  for this variable.
  Gas velocity is indicated by arrows; the fiducial arrow at the lower left
  has length 560~km~s$^{-1}$.
  The length scale bar at the upper left has length 0.5$h^{-1}$~Mpc.
  \label{Fig:final entropy}
  }
\end{figure}

Each merger simulation was followed for approximately 15~Gyr, or about
six sound-crossing times after the first core passage.
Although this time is greater than the age of the Universe for
$\Omega=1$ and $h=0.6$, it is nevertheless insufficient time for the remnants
of either the 1:1 or the 1:3 mergers to come to equilibrium.
This point is illustrated by Figure \ref{Fig:final entropy},
which shows the gas specific entropy and velocity fields
in the $xy$-plane of each run at $t=15$~Gyr.
Although in each case the core of the remnant has lower entropy
than its surroundings, the gas at the margins ($r\ga 5r_s$) continues
to be disturbed by small-scale convective motions as matter accretes
onto the remnant behind the shocks that have been driven out of the
core region.
The offset mergers all show evidence of bulk rotation, with velocities
of $\sim 200-300$~km~s$^{-1}$, but in even
the most offset cases this provides at most a few percent of the
support needed to keep the remnant in hydrostatic equilibrium.
The 1:1 cases are closest to equilibrium, having settled into roughly
spherical distributions within $r\sim R$ of the center of mass.
The 1:3 cases are relatively further from equilibrium, with distinct
streams of material falling in from the right (in the head-on case)
and along the top edge (in the offset cases).
These streams continue to feed high-entropy gas from the smaller cluster
into the remnant as the simulations come to an end.

\begin{figure}[t]
\epsscale{0.5}
\plotone{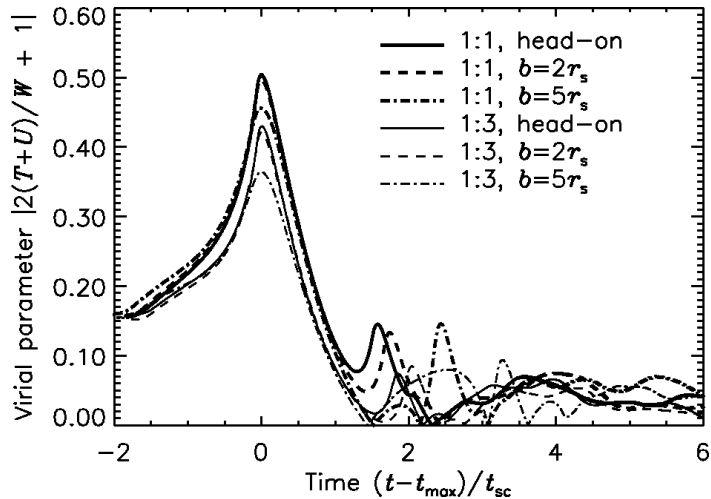}
\caption{
  Virial disequilibrium parameter in the six collision runs.
  Time is offset from the time of peak luminosity $t_{\rm max}$ and
  scaled to the sound-crossing time $t_{\rm sc}$.
  \label{Fig:virial plot}
  }
\end{figure}

Departures from equilibrium can also be described in terms
of total energetics.
In describing the results of gas-only merger simulations,
Ricker (1998) defined a `virial disequilibrium' parameter as
\begin{equation}
V\equiv \left|{2(T+U)\over W} + 1\right|\ ,
\end{equation}
where $T$, $U$, and $W$ are the total kinetic, internal, and potential
energies. When the system is in virial equilibrium, $V=0$, and when it is
not, $V$ is equal to the second time derivative of the moment of inertia,
divided by $W$. Thus a useful equilibrium criterion is to require
$V<\epsilon<<1$ for an extended period (say $\sim t_{\rm sc}$).
In the gas-only mergers, the clusters required between
5.5 and 6 sound-crossing times to reach the level $\epsilon \sim 0.02$,
after which $V$ was roughly constant. In Figure \ref{Fig:virial plot}
we plot $V$ versus time for the six gas-plus-dark matter mergers considered
in this paper. Note that, as in the earlier work, $V$ passes through
a maximum during the first core passage, when the system is furthest from
equilibrium. $V$ then declines steeply but oscillates for several
sound-crossing times as the dark matter continues to pump the gas.
Finally the system begins to tend steadily toward $V=0$ around
$4t_{\rm sc}$ after first core passage. However, by the end of the
simulations only runs C4 and C5 have satisfied the equilibrium
criterion just described.

\begin{figure}[t]
\epsscale{0.75}
\plotone{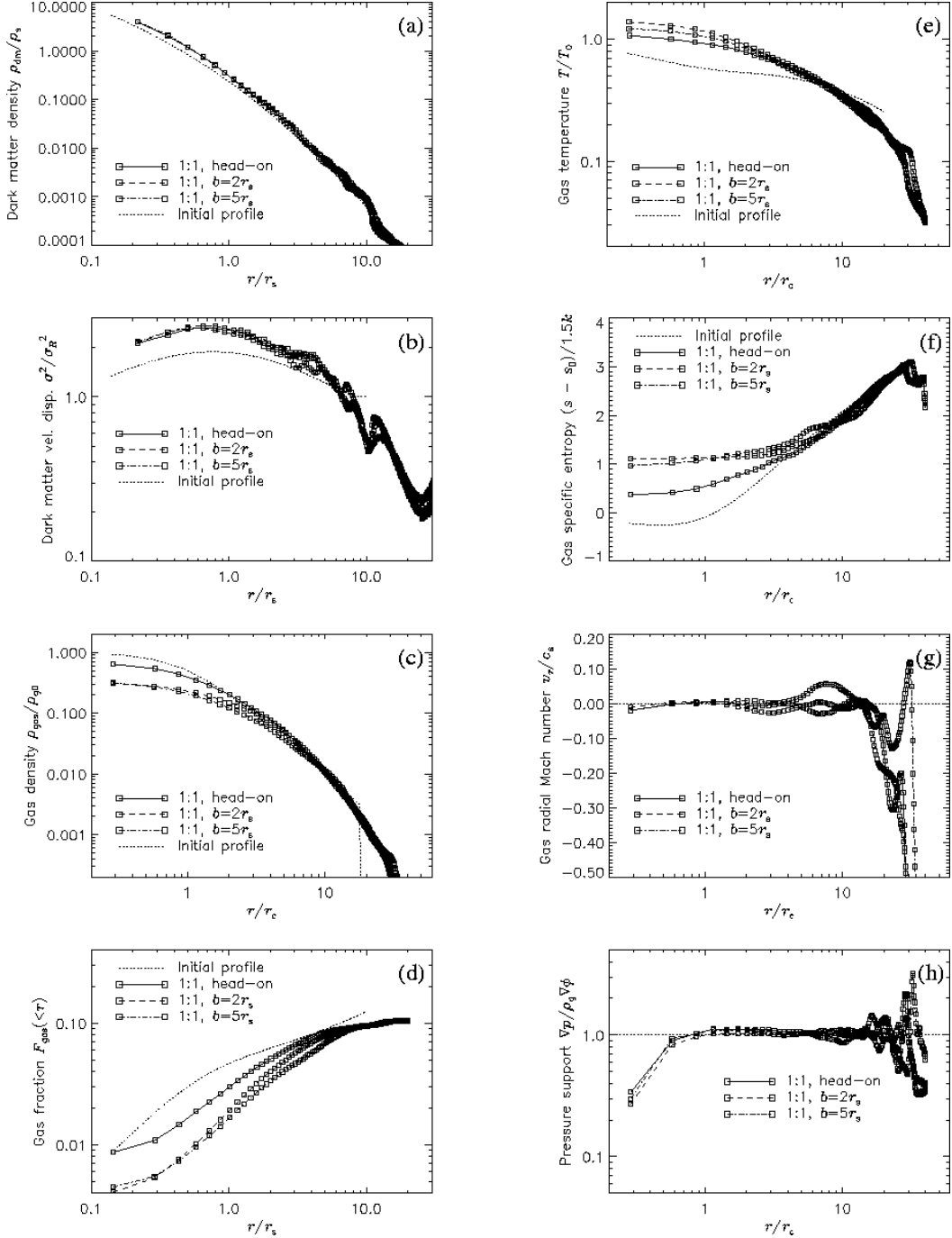}
\caption{
  Average profiles of dark matter and gas quantities at $t=15$~Gyr in the
  1:1 collision runs (C1--C3).
  In all cases except (d), (g), and (h),
  quantities are scaled to their values at the beginning of the simulation.
  (a) Dark matter density.
  (b) Dark matter velocity dispersion.
  (c) Gas density. (d) Enclosed gas fraction. (e) Gas temperature.
  (f) Gas specific entropy. (g) Gas radial Mach number.
  (h) Ratio of pressure force to gravitational force on the gas.
  \label{Fig:1:1 profile plots}
  }
\end{figure}

\begin{figure}[t]
\epsscale{0.75}
\plotone{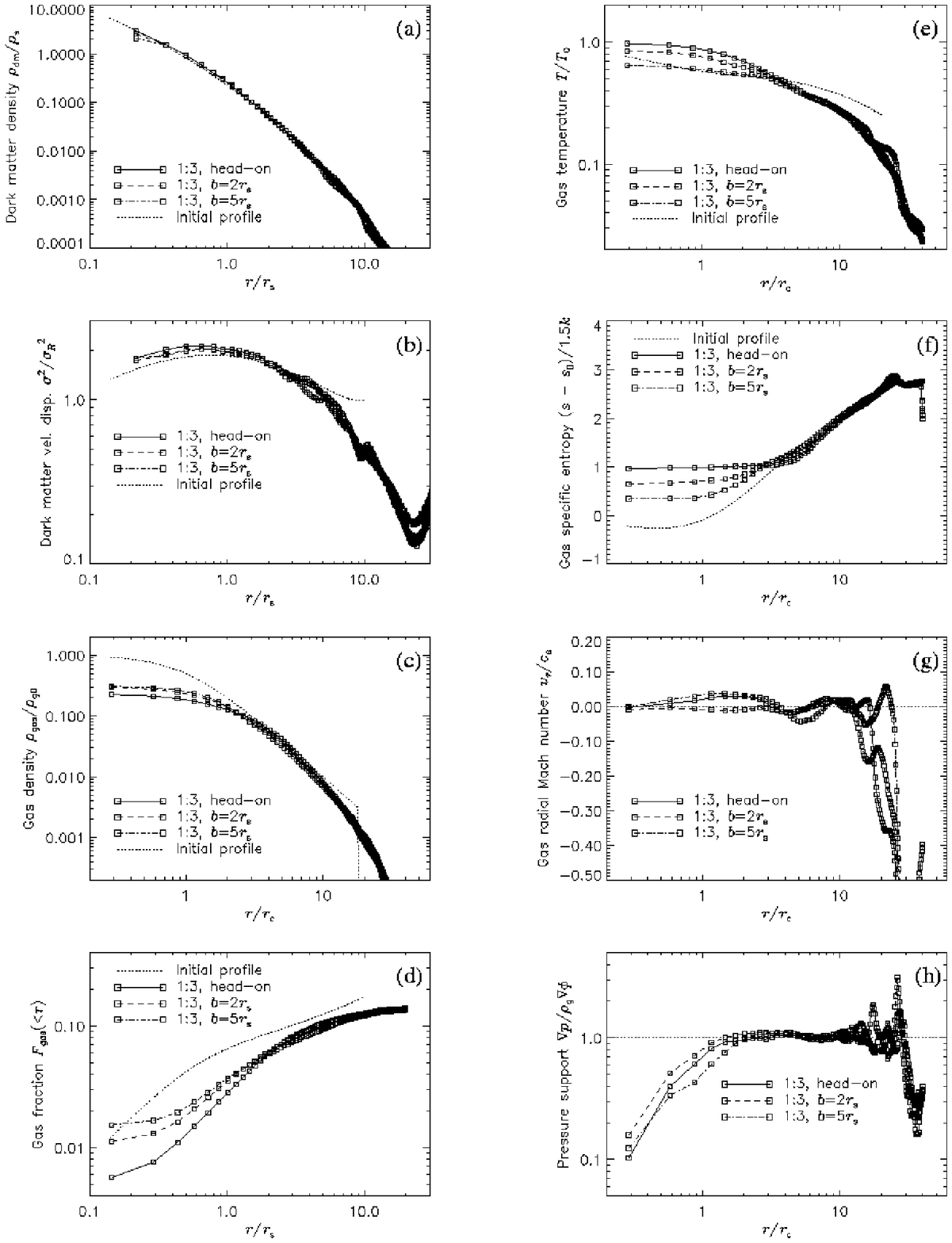}
\caption{
  Average profiles of dark matter and gas quantities at $t=15$~Gyr in the
  1:3 collision runs (C4--C6).
  In all cases except (d), (g), and (h),
  quantities are scaled to their values at the beginning of the simulation.
  (a) Dark matter density.
  (b) Dark matter velocity dispersion.
  (c) Gas density. (d) Enclosed gas fraction. (e) Gas temperature.
  (f) Gas specific entropy. (g) Gas radial Mach number.
  (h) Ratio of pressure force to gravitational force on the gas.
  \label{Fig:1:3 profile plots}
  }
\end{figure}

We plot angle-averaged profiles of various quantities from the merger runs
at $t=15$~Gyr in Figures \ref{Fig:1:1 profile plots} and
\ref{Fig:1:3 profile plots}.
Profiles are generated from the raw simulation data as for the
single-cluster test runs described in \S~\ref{Sec:convergence}.
Although the merger remnant is not in hydrostatic equilibrium, the
average profiles still serve to characterize the effect of the collision
on the structure of the clusters.
As for the single-cluster runs, the profiles here are scaled to the initial
values of $\rho_s$, $r_s$, and so forth, with the exception of the enclosed
gas fraction, the gas radial Mach number, and the pressure support ratio.
For the 1:3 mergers the scaling parameters used are those pertaining to the
more massive cluster (B).

The dark matter density profile
(Figures \ref{Fig:1:1 profile plots}a and \ref{Fig:1:3 profile plots}a)
shows very little change in shape as a result of the collisions.
Of all of the properties of the merger remnant, its shape is also the
least affected by variations in mass ratio and impact parameter.
The normalization ($\sim \rho_s$ since dark matter dominates the potential)
increases in each case because the merger remnant is more massive than
the individual colliding clusters, but the scale radius changes very little.
The asymptotic slope of the profile does not increase as in the
single-cluster test runs.

Figures~\ref{Fig:1:1 profile plots}b and \ref{Fig:1:3 profile plots}b
give the profile of the dark matter velocity dispersion at the
end of the simulations.
At the end of the 1:1 collisions, the central velocity dispersions
are all increased as a result of the deeper potential well of the
merger remnant.
A similar but smaller effect is seen in the 1:3 mergers.
At large radii, the velocity dispersions are all small as a result of
the evaporation of faster moving particles and the outflow-only outer
boundary conditions in the simulations.
A similar effect was seen in the single cluster simulations
(Figure~\ref{Fig:single cluster plots}b).

Although the dark matter density profiles are not strongly affected by the mergers,
the gas density profiles
(Figures \ref{Fig:1:1 profile plots}c and \ref{Fig:1:3 profile plots}c)
change substantially as a result of the mergers.
The central gas density in each case actually shows a substantial decrease
from its initial value, dropping to 30--70\% of its initial value for the
1:1 runs and to 20--30\% of its initial value for the 1:3 runs.
Using the half-density radius as an estimate for the core radius, we
find that $r_c$ increases by 20--25\%.
The final value of $\rho_{g0}$ shows a significant dependence on impact
parameter.
For the equal-mass collisions, the final gas density is about twice as
high for head-on mergers as for the most offset mergers.
For the 1:3 collisions, the variation is only 50\%, but the sense is
reversed, with head-on collisions resulting in a lower central gas density.
In all cases we see a steepening of the asymptotic slope from
$r^{-2}$ to $r^{-3}$, as in the single-cluster runs, because of the
initial cutoff at $r=R$ combined with the outflow boundary conditions.

The flattening of the gas density profile occurs because
the shocks generated by the merger add entropy to the system
and increase the buoyancy of the ICM in the relatively constant
dark matter potential.
The resulting profile for the enclosed gas fraction
(Figures \ref{Fig:1:1 profile plots}d and \ref{Fig:1:3 profile plots}d)
typically begins below the initial value of $\sim 1\%$ for radii
comparable to the zone spacing, then increases roughly as $r^{0.8-1.0}$
and almost meets the initial profile at $r\sim R/2$ before flattening out to
an asymptotic value slightly below the initial value.
The change in curvature at small radii is also seen in the single-cluster
runs and is a numerical effect due to the influence of particle-mesh
force smoothing on the cuspy dark matter profile.
Likewise, the drop from the initial profile between $r\sim R/2$ and
$r\sim R$ is a numerical effect due to the expansion of material beyond
the imposed initial density cutoff.
The drop in the asymptotic value of the gas fraction occurs because
more gas than dark matter is lost through the outflow boundary --
$\sim 30\%$ versus $\sim 16\%$ -- as shocks tend to eject collisional
material while leaving the collisionless dark matter alone.
The addition of an external confining pressure would most likely
allow the asymptotic gas density slope to be flatter but leave the
inner core unchanged.

This is apparent from the gas temperature profiles
(Figures \ref{Fig:1:1 profile plots}e and \ref{Fig:1:3 profile plots}e),
which show an inner region whose behavior depends on impact parameter and
mass ratio and an outer region that is roughly independent of impact
parameter and corresponds to the single-cluster result.
The boundary between these two regions occurs at $r\sim 3r_c$ in each
of the runs and is $\sim 20\%$ hotter relative to $T_0$ in the 1:1 runs
than in the 1:3 runs.
In all cases the final temperature profile declines more slowly at the
very center than does the initial profile.
Offset mergers display opposite trends with increasing
$b$ for the two mass ratios: for 1:1 mergers, increasing impact parameter
increases the final temperature over the head-on case, while for 1:3 mergers
it decreases the temperature, reproducing the initial profile in the
most offset 1:3 case.
This behavior occurs because of the increase in core entropy associated
with mixing of shocked gas from the outer ICM into the core material,
as discussed in \S~\ref{Sec:pressure peaks}.
The specific entropy profile
(Figures \ref{Fig:1:1 profile plots}f and \ref{Fig:1:3 profile plots}f)
also demonstrates a parameter-independent
outer region and parameter-dependent core, with the central entropy
increase being $\sim 1.5k$ over the initial central value in the hottest
runs.
Note that the magnitude of this increase is roughly the same for the two
mass ratios.

Although the gas has not reached virial equilibrium in the simulations
by $t=15$~Gyr, it is relatively quiescent.
Figures \ref{Fig:1:1 profile plots}g and \ref{Fig:1:3 profile plots}g
show the radial Mach number profiles for
each run. Within $r=10r_c$ the radial Mach number is less than 0.06 in
magnitude, indicating that radial motions near the center are fairly
subsonic at this point. Outside $r=10r_c$, the magnitude of the Mach
number increases rapidly, reflecting the continued accretion of gas
that was previously ejected from the center by the merger shocks.
The gas within the initial cutoff radius is also quite close to
hydrostatic equilibrium in angle average.
The angle-averaged ratio of the pressure force to gravitational force
per unit mass is shown in
Figures \ref{Fig:1:1 profile plots}h and \ref{Fig:1:3 profile plots}h.
Except for the innermost few zones where the small number of zones in
the radial bins and truncation error in the gradient calculation
dominate this ratio, it is within 15\% of unity for $r<10r_c$.


\subsection{Luminosity and Temperature Evolution}
\label{Sec:luminosity-temperature}

In the simplest self-similar picture of their assembly,
clusters undergo spherical collapse and virialization and then radiate via
thermal bremsstrahlung (Kaiser 1986).
This model fails to predict the observed X-ray luminosity-temperature
relationship: the predicted relation is $L_X\propto T_X^2$, whereas
the observed relation is closer to $L_X\propto T_X^3$
(Mushotzky 1984).
Cooling flows have larger luminosities and lower average temperatures
than average and are known to steepen this relation and increase its
dispersion (Fabian et al.\ 1994).
However, removing them from cluster samples in various ways
does not eliminate the discrepancy (Markevitch 1998; Allen \& Fabian 1999;
Arnaud \& Evrard 1999).
Various solutions to this problem have been proposed.
Most recently these have focused on raising the entropy of the ICM
via nongravitational heating, though Bryan (2000) has recently suggested
that galaxy formation may do the same thing by removing low entropy gas.
Preheating models increase the entropy of the gas before it falls into
cluster potentials
(Kaiser 1991; Evrard \& Henry 1991).
These models require substantial amounts of heating:
$\sim 1$~keV/baryon appears to be needed (Lloyd-Davies, Ponman,
\& Cannon 2000).

Note that the luminosity-temperature relation used
to scale the initial conditions for our simulations
(eq.~[\ref{Eqn:observed L-T}]) is
not retained as the mergers progress.
Comparing $T_X^{a_L}$ to $L_X$ as time passes in run C1, we find that
immediately following the initial brightness peak, the ratio of these
two quantities jumps to twice its initial value, fluctuates here for a bit,
increases briefly to almost six times its initial value during the second
core passage, then settles down to about three times its initial value,
where it remains roughly constant for the remainder of the calculation.
The behavior of the other runs is similar.
Thus the trajectory of the system in the $L_X-T_X$ plane is roughly
parallel to the observed statistical relation both before and after the
merger, but during the merger the system undergoes a shift to a higher
temperature normalization.

The normalization of the $L_X-T_X$ relation is complicated by the
presence of two clusters within the computational volume.
In this context we note that luminosity is an `extensive' quantity,
which must be divided by the number of clusters one is counting,
while temperature is an `intensive' quantity, which does not scale with
the number of clusters.
For equal-mass mergers,
the cluster cores merge to form a single core (to within the spatial
resolution of our grid) approximately 5--7~Gyr after the initial core
passage.
Long before this point the clusters would be considered a single
cluster with `bimodal' structure if seen along the $z$-axis, because
the twin brightness peaks lie within an Abell radius of each other and
are gravitationally bound.
Moreover, from most lines of sight the cluster would appear to be
unimodal by at most 3~Gyr after initial core passage.
Indeed, the head-on collision seen along the collision axis appears to
be unimodal throughout the calculation.
Comparing the initial and final states of the system is less problematic.
Initially we have two clusters following the imposed $L_X-T_X$ relation;
at the end, we have one cluster with temperature 1.4 times the initial
value and luminosity between 80\% and 160\% of the initial value of the
individual clusters.
Hence the final ratio of $T_X^{a_L}$ to $L_X$ is in the range
1.5 (run C1) to 3 (run C3).

Alternatively, one may say that the final temperature is $\sim 1.2-1.5\times$
the value expected given the final luminosity and
the initial $L_X-T_X$ relation.
Let us define
\begin{equation}
f_{LT} \equiv {T_X^{\rm final}/T_X^{\rm initial}\over
	\left(L_X^{\rm final}/L_X^{\rm initial}\right)^{1/{a_L}}}\ .
\end{equation}
In the highest-resolution single-cluster run (S7) we see
$f_{LT} \sim 1.04$ due to numerical effects.
Nearly all of the X-ray emission arises within a few scale radii of
the center of the cluster, so the increase in run S7 (and presumably a
comparable portion of the increase in the merger runs) is probably
due to spatial resolution.
The lack of an external confining pressure
(such as might be expected due to continuously infalling material
from the low-density intergalactic medium)
allows material from the center of the cluster to expand more than it might
otherwise, causing unwanted adiabatic cooling.
Although this changes $f_{LT}$, adiabatic changes yield
$L_X\propto T_X^2$, so cooling due to the outflow boundaries should
{\sl decrease} this ratio, not increase it:
\begin{equation}
f_{LT}|_{\rm adiabatic} \approx \left({T_X^{\rm final}\over
	T_X^{\rm initial}}\right)^{1-2/{a_L}} \approx
	\left({T_X^{\rm final}\over
	T_X^{\rm initial}}\right)^{0.24}\ .
\label{eq:fLTadiab}
\end{equation}
Shocks might have a much larger effect on $f_{LT}$ because they produce
less of a density jump than a temperature jump.
For shocks of arbitrary strength, the inverse of the shock compression is
given by
(Markevitch, Sarazin, \& Vikhlinin 1999):
\begin{equation}
\frac{\rho_1}{\rho_2} =
\left[\frac{1}{4}\left(\frac{\gamma+1}{\gamma-1}\right)^2
\left(\frac{T_2}{T_1}-1\right)^2 +\frac{T_2}{T_1}\right]^{1/2}
-\frac{1}{2}\frac{\gamma+1}{\gamma-1}\left(\frac{T_2}{T_1}-1\right) ,
\end{equation}
where $\rho_1$ ($\rho_2$) and $T_1$ ($T_2$) are the preshock (postshock)
density and temperature, respectively, and $\gamma$ is the adiabatic
index.
For $\gamma = 5/3$, this gives
\begin{eqnarray}
f_{LT}|_{\rm shock} & \approx &
\left( \frac{T_X^{\rm final}}{T_X^{\rm initial}} \right)^{1-1/2{a_L}}
\left\{ \left[ 4 \left( \frac{T_X^{\rm final}}{T_X^{\rm initial}} -1\right)^2
+ \frac{T_X^{\rm final}}{T_X^{\rm initial}} \right]^{1/2}
-2 \left( \frac{T_X^{\rm final}}{T_X^{\rm initial}} -1 \right)
\right\}^{1 / a_L} \nonumber \\
& \approx &
\left( \frac{T_X^{\rm final}}{T_X^{\rm initial}} \right)^{0.81}
\left\{ \left[ 4 \left( \frac{T_X^{\rm final}}{T_X^{\rm initial}} -1\right)^2
+ \frac{T_X^{\rm final}}{T_X^{\rm initial}} \right]^{1/2}
-2 \left( \frac{T_X^{\rm final}}{T_X^{\rm initial}} -1 \right)
\right\}^{0.38} \, .
\label{eq:fLTshock}
\end{eqnarray}
The average temperature increases by about 40\% during the equal-mass
merger runs, so the value of $f_{LT}$ lies between 1.2 and 1.5.
In the 1:3 cases,
the less massive cluster contributes only about 10\% as much luminosity
as the more massive cluster at the start of the calculation, so
in this case $f_{LT} \approx 1.4$, consistent with the range of values
in the 1:1 collisions.
On the other hand, equations~(\ref{eq:fLTadiab}) and (\ref{eq:fLTshock})
give
$f_{LT}|_{\rm adiabatic} \approx 1.08$ and
$f_{LT}|_{\rm shock} \approx 1.10$.
Thus, the temperature increase is larger (or the X-ray luminosity increase
is smaller) than expected from either shocks or adiabatic compression.
This probably indicates that the luminosity increase is limited by the finite
numerical resolution in the cores of the clusters and would be larger
if the calculations were better resolved.

\begin{figure}[t]
\epsscale{0.75}
\plotone{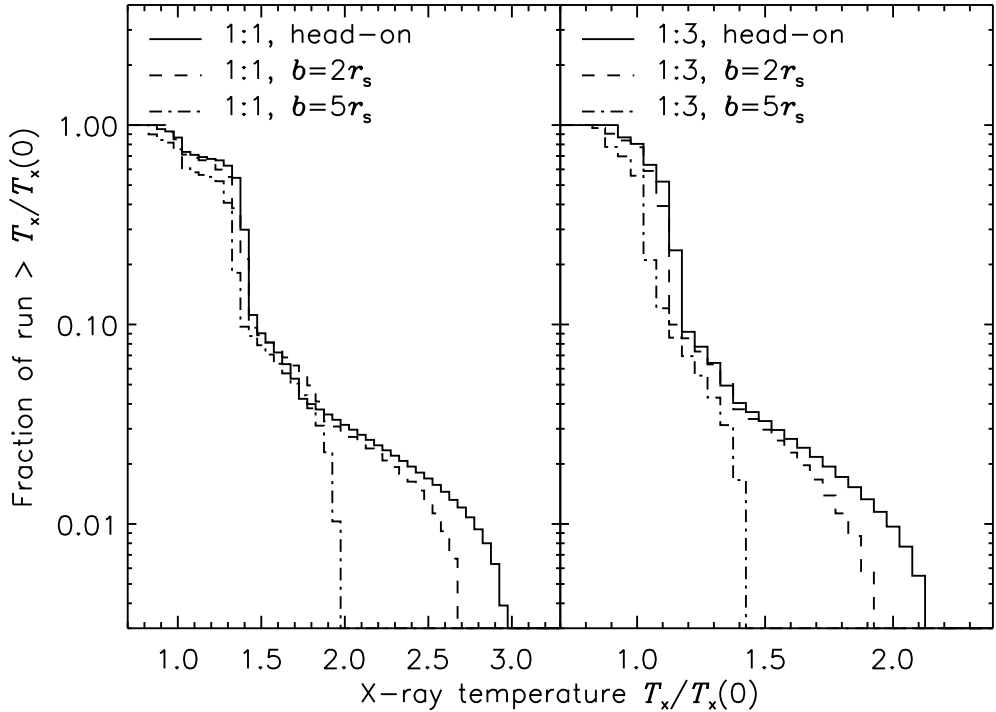}
\caption{
  Cumulative distribution of X-ray temperatures $T_X$ for each run
  for times between $-2t_{\rm sc}$
  and $+6t_{\rm sc}$ of the time of maximum luminosity. X-ray temperatures are
  expressed in terms of the initial temperature, $T_X(0)$, and accumulated in
  bins of width 0.05. Edges are visible at the locations of the initial and
  final temperatures, and changes in slope are visible at the maximum
  temperatures
  reached during the first and second core passages. For approximately
  8\% of each run in the 1:1 cases, and 3\% in the 1:3 cases,
  the system has a temperature greater than 1.5 times its initial value.
  \label{Fig:temphist}
  }
\end{figure}

We close this section with the following note.
The observation of massive clusters at high redshift (e.g., MS1054: Donahue et
al.\ 1998) is often taken as strong evidence for low $\Omega_0$ because of the
exponential sensitivity of the cluster mass function to $\Omega_0$ (e.g.,
Oukbir \& Blanchard 1992).  However, in deriving such constraints one directly
measures an X-ray temperature, then uses the mass-temperature relation to find
the mass.  Our results show that this procedure is dangerous:  merger shocks
produce short-lived, but significant, increases in the X-ray temperature of a
cluster.  For a few percent of its life, a merging cluster appears much hotter,
and thus more massive, than it really is (Figure \ref{Fig:temphist}).
Furthermore, because this period
corresponds to a time of significantly enhanced luminosity, merging clusters
are also easier to detect.  We note, for example, that MS1054 now shows
evidence of substructure and a lower intrinsic temperature than first reported,
suggesting that its apparent high temperature is due to an ongoing merger
(Neumann \& Arnaud 2000).
As another example, 
G\'omez, Hughes, and Birkinshaw (2000) have recently suggested that
Abell~665, one of the richest, hottest, and most X-ray luminous clusters
at redshifts $z \la 0.2$, is currently undergoing a merger and is
at the epoch when the cluster cores collide.
This could explain its very high luminosity and temperature.
We are using our results with a merger tree code to
estimate the effect of merger shocks on the significance of distant, massive
cluster constraints on $\Omega_0$ and will address this question in a
subsequent paper (Randall, Sarazin, \& Ricker 2001, in preparation).

\subsection{Persistence of Gas Cores and Turbulence}
\label{Sec:pressure peaks}

A cluster is defined by the presence of a distinct potential
well within which dark matter, gas, and galaxies are confined.
Mergers produce dramatic fluctuations in these potentials,
leading eventually to the combination of the merging clusters into a single
remnant.
Some merging clusters in which the mergers appear to be well-advanced show
a double core structure at their centers (e.g.,
Abell~2065; Markevitch et al.\ 1999),
with two X-ray surface brightness peaks, indicating that the cores of the
individual cluster potentials are still intact.
If the original subclusters have cooling flows at their centers,
the X-ray surface brightness of the surviving peaks in the subcluster
potentials can be particularly large.

Cluster cores lose their distinct identity in several ways.
As with galaxies that pass too close to a cluster's center, tidal stripping
can remove material from a shallower cluster potential and add it to a deeper
one. In addition, violent relaxation (for collisionless matter) and
ram-pressure stripping (for diffuse gas) convert the bulk kinetic energy of a
cluster into dispersive or thermal energy, enabling material to become unbound
from its own potential, while remaining bound within the combined potential
of the material from both clusters. Dynamical friction offers another means
for converting organized bulk velocities to dispersive ones, but the timescale
on which
it operates is typically large compared to the timescales associated with
these other mechanisms.

Our results shed some light on the mechanism by which ram pressure affects
the survival of cluster cores in mergers.  Rather than ablating gas from the
edges of a core, ram pressure disrupts cores by pushing the gas as a unit
away from the center of the potential well in which it originates, causing
it to become convectively unstable. The resulting convective motions destroy
the spherical symmetry of the core and mix the core gas with higher-entropy
shocked gas in its surroundings. This also appears to be an important mechanism
by which mergers initiate turbulent motions.

\begin{figure}[t]
\epsscale{0.75}
\plotone{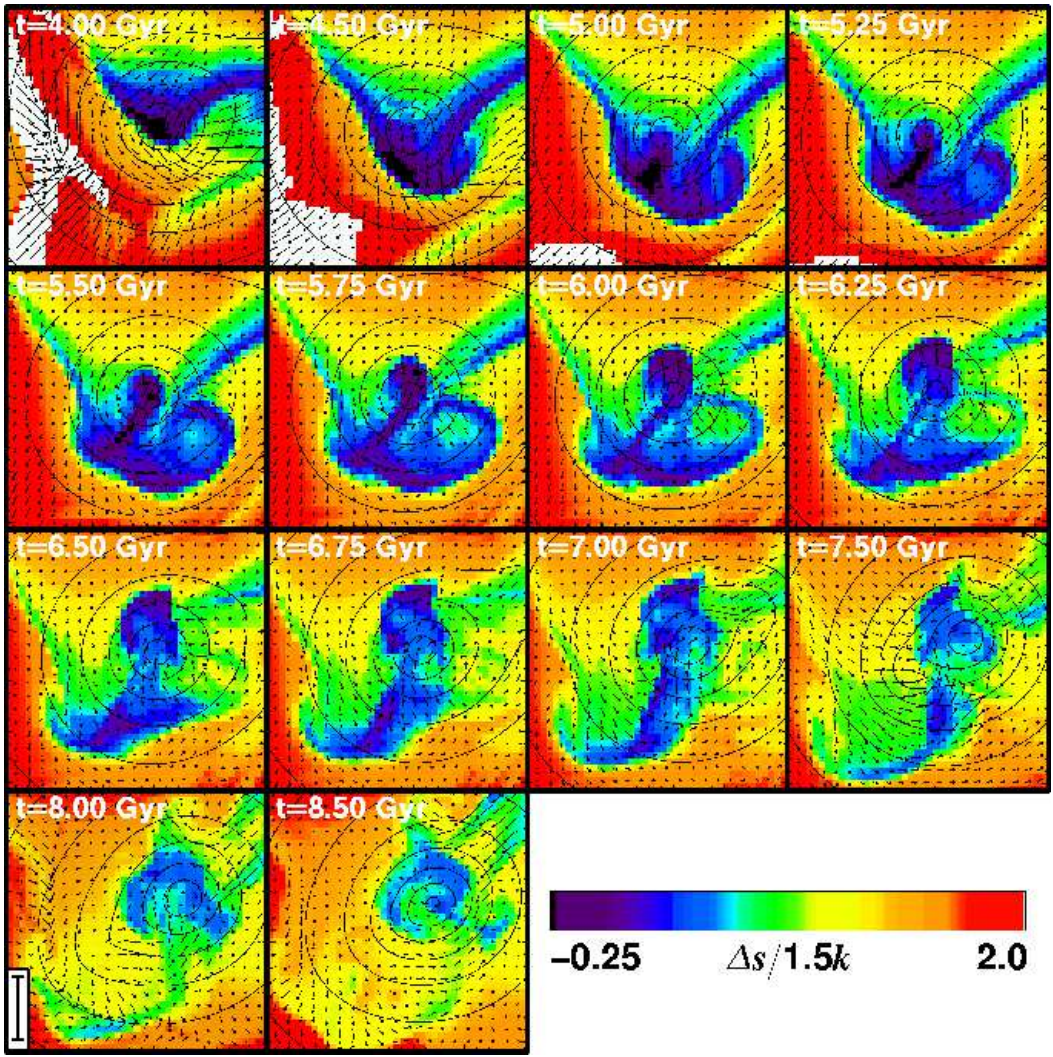}
\caption{
  A merger-driven convective plume. The plots show
  specific entropy, measured relative to the initial central value of the more
  massive cluster, in the collision plane of run C5.  The colormap is chosen to
  accentuate low-entropy regions, so the shocked region in the first four
  frames, where the colormap is saturated, appears white.  Contours show the
  gravitational potential, while arrows show the gas velocity field.  The
  fiducial bar at lower left has a length of $250h^{-1}$~kpc.
  \label{Fig:convective plume}
  }
\end{figure}

Figure \ref{Fig:convective plume} shows an example of this process using the
specific entropy field.
This example begins just after the first core interaction of run C5 (the less
massive cluster core has exited to the right at 4.00 Gyr).  The diffusivity of
the gas is low, so the core interaction pushes low-entropy gas out of the more
massive core, where it finds itself out of convective equilibrium.  Beginning
around 5.00 Gyr this gas flows in a plume back toward the more massive
potential center.  A large eddy also forms in the wake of the less massive
cluster; its remnants persist until 7.00 Gyr.  The plume overshoots the
potential center and begins to decelerate and spread along equipotential lines.
It reaches maximum overshoot around 7.00 Gyr, when it is met by high-entropy
gas from the returning less massive core.  The remaining gas bound to this core
has higher entropy and, stripped from its potential well, is forced to flow
around the plume.  As the dark matter associated with the less massive cluster
passes through the more massive potential well, it drives a new shock through
the low-entropy gas from the more massive cluster that did not flow into the
plume; this shock exits to the lower left after 8.50 Gyr.  The
remnants of the plume mix with the newly reshocked gas to form the core of the
merger remnant, with higher entropy than was present in the original clusters.

\begin{figure}[t]
\epsscale{0.5}
\plotone{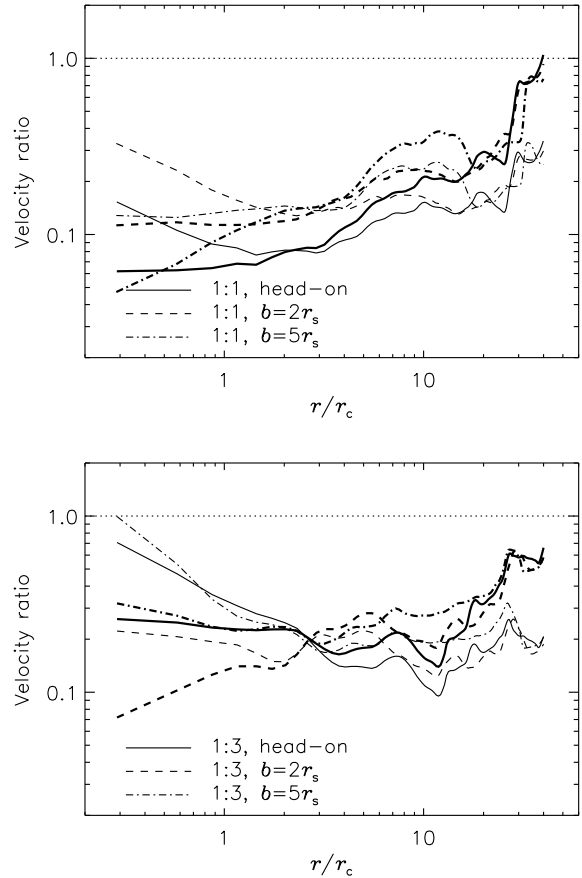}
\caption{
  rms gas velocity in runs C1-C6 at $t\approx15$~Gyr as a fraction of the
  Keplerian circular velocity (thin lines) and the angle-averaged sound speed
  (thick lines). rms velocity is defined here as
  $\left(\sigma_r^2+\sigma_\theta^2+\sigma_\varphi^2\right)^{1/2}$,
  with the component rms velocities $\sigma_r$,
  $\sigma_\theta$, and $\sigma_\varphi$ accumulated in radial bins about the
  center of mass.
  \label{Fig:rms gas velocities}
  }
\end{figure}

The turbulent velocity field created by a merger lasts for several
sound-crossing
times. In Figure \ref{Fig:rms gas velocities} we plot
the rms gas velocity profile in each of the collision runs at $t\approx15$~Gyr
as a fraction of the average sound speed and the circular (Keplerian)
velocity. Throughout the merger remnant,
turbulent velocities are subsonic: for $r\la 20r_c$ they are
typically $10-20$\% of the sound speed in the 1:1 cases, while the rms Mach
number is about twice as large in the 1:3 cases.
Outside this radius the rms velocity rises to between 60\% and 100\% of the
sound speed, with the increase much steeper in the 1:1 cases.
Throughout
most of the cluster, turbulence supplies $\sim 5-10$\% of the hydrostatic
support. In certain cases this contribution appears to be larger within the
innermost $r_c$ or so of the core, though it is likely that here the rms
velocity is overestimated due to the relatively small number of zones
contributing to radial velocity averages.

Our results support the finding of others (e.g., Roettiger, Burns,
\& Loken 1996; Norman \& Bryan 1998) that,
following a merger, the intracluster medium becomes turbulent,
and that turbulent pressure and bulk rotation provide less than
20\% of the support needed for hydrostatic equilibrium. The picture which
emerges from our calculations is that this turbulence is initiated by
convective instability and, to a lesser extent, by baroclinic vorticity
generation at oblique shocks. It is sustained for several Gyr by pumping
due to the oscillation of the dark matter, which relaxes more slowly than
the gas. While Roettiger et al.\ (1996) also argue for this
mechanism using isolated merger simulations, Norman and Bryan (1998) find
higher rms velocities in a cluster simulation involving multiple mergers
and accretion, and they argue that minor mergers are the primary pumping
mechanism. Our results suggest that individual major mergers
and cumulative minor mergers induce comparable amounts of turbulence,
since the turbulent pressure in our simulations is $\sim 1/2 - 1/3$
that of Norman \& Bryan's result even several Gyr after the collision.

To date the grid-based merger simulations used to argue for
turbulence in clusters have reached spatial resolutions on the order of 10~kpc.
In fully developed turbulence, the range of eddy length scales is of order
\begin{equation}
\label{Eqn:turbulence reynolds number}
{\ell_0\over\ell_d} \sim (Re)^{3/4}\ ,
\end{equation}
where $\ell_0$ is the integral scale and $\ell_d$ is the dissipation scale
(Richardson 1922; Kolmogoroff 1941).
Thus, directly simulating three-dimensional turbulent flow with a
hydrodynamic code requires at least of order $(Re)^{9/4}$ zones.
For many astrophysical applications, this is impossible -- for example,
in stellar convective layers $Re\sim10^{14}$.
However, as we have already argued, the Reynolds number of the intracluster
medium during mergers is of order $10^3$, and turbulence, if it is present
at all, operates over a much smaller range of length scales.
We therefore have some hope of performing direct numerical simulations of
merger-driven turbulence within the very near future.
By comparison, the simulations described in this paper achieve a spatial
dynamic range of $\sim 200$.

Several approximate methods for numerically treating turbulence exist
(see Canuto 1994 for a partial review).
Among these methods, large-eddy simulation (LES) has achieved some
success in reproducing characteristics of turbulence in laboratory
experiments (Canuto 1997).
LES attempts to simulate directly only the largest-scale eddies in
a turbulent flow, since these depend on the boundary conditions and
geometry of the flow at the integral scale, and it uses a subgrid model
to handle the (presumably universal and geometry-independent)
turbulent dissipation on unresolved scales.
Most grid-based astrophysical simulation algorithms are, on some level, LES,
but they do not use subgrid turbulence models, instead relying on
numerical dissipation at the zone scale.
This numerical dissipation arises due to the truncation error of each
individual method and bears no relation to the real dissipation mechanism.
For most methods this practice should yield an incorrect energy
dissipation spectrum, since it links large- and small-scale eddies in an
unphysical way (Canuto 1997).
However, it has been argued by Sytine et al.\ (2000) that the
piecewise-parabolic method, which we employ in this paper, accurately
reproduces the energy and enstrophy power spectra produced by solving
the full Navier-Stokes equations in direct simulations of compressible
turbulence, because it is much less dissipative than other techniques
and because its dissipation is confined to narrow regions near flow
discontinuities.
Estimates of the dissipation in PPM for some simple flow fields
(Porter \& Woodward 1994) suggest that convective features spanned
by 32 or more zones are well-converged.
This is consistent with recent results of Calder et al.\ (2001) for
single-mode Rayleigh-Taylor instabilities.
Thus, for example, the convective plume shown in Figure
\ref{Fig:convective plume}, which measures approximately
32$\times$10 zones at maximum extent, is probably just sufficiently resolved
to yield the correct growth rate.
Further high-resolution calculations using adaptive mesh
refinement will be useful in verifying this result (Ricker et al.\ 2001).


\section{Summary}
\label{Sec:conclusions}

We have presented results from a controlled parameter study of off-axis
mergers between clusters of galaxies. We have extended previous work by
using the NFW density profile and nonisothermal temperature profiles,
systematically controlling merger parameters and justifying their scaling,
carefully studying numerical effects and convergence requirements,
and examining the survival of pressure peaks and the mechanism for the
onset of turbulence in mergers. We have also presented results on the variation
of relaxation times, merger remnant properties, and luminosity and temperature
peaks with merger parameters. The calculations presented here do not include
radiative cooling or magnetic fields. We anticipate adding these
physical processes in future calculations (particularly cooling, which
appears to play a very important dynamical role in many observed clusters).

The morphological changes, relative velocities, and temperature jumps
we observe agree well with previous studies of collisions between
clusters modeled using the King profile (e.g., Roettiger et al.\ 1998).
In particular, velocities of $\sim 2000$~km~s$^{-1}$ are to be expected,
even in quite offset mergers, for cosmologically reasonable initial conditions.
The peak temperature occurs for a period $\la t_{\rm sc}/2$ at first core
passage and is a factor $\sim 2-3$ greater than the initial temperature.
We observe a larger jump in X-ray luminosity ($\sim 4-10\times$)
than previous studies including dark matter,
and we argue that this increase is most likely a lower limit due to our
spatial resolution. Second core passage, with corresponding secondary
luminosity and temperature jumps, typically occurs $1.5-2.5t_{\rm sc}$ after
the first. In the most offset 1:3 case, this secondary peak is delayed by
an additional sound-crossing time. We emphasize that luminosity and temperature
jumps due to mergers may have an important bearing on constraints on $\Omega$
derived from the observation of hot clusters at high redshift. We will address
this issue further in a subsequent paper.

Shocks play an important dissipative role in mergers, but they are relatively
weak in the highest-density regions. As a result they do not directly raise the
entropy of the cluster cores. Instead, shocks create entropy
in the outer parts of the clusters, and this high-entropy gas is mixed with
the core gas during later stages of the merger. Mixing is initiated by
ram pressure: the core gas is displaced from its potential center and becomes
convectively unstable. The resulting convective plumes initiate large-scale
turbulent motions with eddy sizes up to several 100~kpc. This turbulence
is pumped by oscillations in the gravitational potential, which in turn are
driven by the more slowly relaxing collisionless dark matter.
Even after nearly a Hubble time these motions persist as subsonic
turbulence in the cluster cores, providing $5-10\%$ of the support against
gravity.
The dark matter oscillations are also reflected in the extremely long
time following a merger required for the remnant to come to virial equilibrium.
Because of the increase in core entropy, if a constant-density
gas core is initially present, it is not destroyed by a merger,
even if the dark matter density profile has a central cusp.



\acknowledgments

This research has been supported in part by NASA Astrophysical
Theory Program grant NAG 5-3057 and
{\it Chandra} Award Numbers GO0-1019X, GO0-1141X, GO0-1158X,
and GO0-1173X
at the University of Virginia,
and by
the ASCI Flash Center at the University of Chicago under DOE
contract B341495. Calculations were performed using the computational
resources of the Pittsburgh Supercomputing Center and the San Diego
Supercomputer Center.
P.~M.~R.\ would like to acknowledge useful conversations with
D.~Lamb, M.~C.~Miller, J.~Stone, and E.~Ostriker.





\end{document}